\documentclass[11 pt] {article}
\usepackage{graphicx}
\usepackage{latexsym}
\usepackage{amsfonts}
\usepackage{amsmath}
\usepackage{enumerate}

\newcommand\blackslug{\hbox{\hskip 1pt \vrule width 4pt height 8pt depth 1.5pt
        \hskip 1pt}}
\newcommand\bbox{\hfill \quad \blackslug \medbreak}
\def\d{\hbox{-}}
\def\c{\hbox{-}\cdots\hbox{-}}

\DeclareMathOperator{\Mark}{Mark}
\DeclareMathOperator{\Move}{Move}
\DeclareMathOperator{\Unmark}{Unmark}
\DeclareMathOperator{\Explore}{Explore}
\DeclareMathOperator{\State}{State}

\newtheorem{theorem}{}[section]
\newtheorem{lemma}[theorem]{}

\newcounter{claim}

\newcommand{\Proof}{\setcounter{claim}{0}\noindent{\bf Proof.}\ \ }

\title{Coloring perfect graphs with no balanced skew-partitions}
\author{Maria Chudnovsky\thanks{Columbia University, New
    York. Partially supported by NSF grants DMS-1001091 and
    IIS-1117631.}~, Nicolas Trotignon\thanks{CNRS, LIP, ENS Lyon,
    INRIA, Universit\'e de Lyon.}~, Th\'eophile Trunck\thanks{ENS Lyon,
    LIP, INRIA, Universit\'e de Lyon.}\\ and Kristina Vu\v
  skovi\'c\thanks{School of Computing, University of Leeds, and
    Faculty of Computer Science (RAF), Union University, Belgrade,
    Serbia.  Partially supported by EPSRC grant EP/H021426/1 and
    Serbian Ministry of Education and Science projects 174033 and
    III44006.  \newline The second and third authors are partially
    supported by \emph{Agence Nationale de la Recherche} under
    reference \textsc{anr 10 jcjc 0204 01}. The three last authors are
    partially supported by PHC Pavle Savi\'c grant 2010-2011, jointly
    awarded by EGIDE, an agency of the French Minist\`ere des Affaires
    \'etrang\`eres et europ\'eennes, and Serbian Ministry of Education
    and Science.}}

\begin{document}
\maketitle

\begin{abstract}
  We present an $O(n^5)$ algorithm that computes a maximum stable set
  of any perfect graph with no balanced skew-partition.  We present
  $O(n^7)$ time algorithm that colors them.
\end{abstract}

\section{Introduction}

A graph $G$ is \emph{perfect} if every induced subgraph $G'$ of $G$
satisfies $\chi(G') = \omega(G')$.  In the 1980's, Gr\"ostchel,
Lov\'asz, and Schrijver~\cite{gls:color} described a polynomial time
algorithm that colors any perfect graph.  A graph is \emph{Berge} if
none of its induced subgraphs, and none of the induced subgraphs of its
complement, is an odd chordless cycle on at least five vertices.
Berge~\cite{berge:61} conjectured in the 1960s that a graph is Berge
if and only if it is perfect.  This was proved in 2002 by Chudnovsky,
Robertson, Seymour and Thomas~\cite{CRST}.  Their proof relies on a
\emph{decomposition theorem}: every Berge graph is either in some
simple basic class, or has some kind of decomposition.  In 2002,
Chudnovsky, Cornu\'ejols, Liu, Seymour and Vu\v
skovi\'c~\cite{chudnovsky.c.l.s.v:reco} described a polynomial time
algorithm that decides whether any input graph is Berge.  The method
used in \cite{gls:color} to color perfect graphs (or equivalently
by~\cite{CRST}, Berge graphs) is based on the ellipsoid method,
and so far no purely combinatorial method is known.  In particular, it
is not known whether the decomposition theorem from~\cite{CRST} may be
used to color Berge graphs in polynomial time.

This question contains several potentially easier questions.  Since
the decomposition theorem has several outcomes, one may wonder
separately for each of them whether it is helpful for coloring.  The
basic graphs are all easily colorable, so the problem is with the
decompositions.  One of them, namely the balanced skew-partition,
seems to be hopeless.  The other ones (namely the $2$-join, the
complement $2$-join and the homogeneous pair) seem to be more useful
for coloring and we now explain the first step in this direction.
Chudnovsky~\cite{thesis,trigraphs} proved a decomposition theorem for
Berge graphs that is more precise than the theorem from~\cite{CRST}.
Based on this theorem, Trotignon~\cite{nicolas:bsp} proved an even
more precise decomposition theorem, that was used by Trotignon and
Vu\v skovi\'c~\cite{nicolas.kristina:2-join} to devise a polynomial
algorithm that colors Berge graphs with no balanced skew-partition,
homogeneous pair nor complement $2$-join.  This algorithm focuses on
the $2$-join decompositions.  Here, we strengthen this result by
constructing a polynomial time algorithm that colors Berge graphs with
no balanced skew-partition.

Our algorithm is based directly on \cite{thesis,trigraphs}, and a few
results from~\cite{nicolas:bsp} and \cite{nicolas.kristina:2-join} are
used.  It should be pointed out that the method presented here is
significantly simpler and shorter than \cite{nicolas.kristina:2-join},
while proving a more general result.  This improvement is mainly due
to the use of \emph{trigraphs}, that are graphs where some edges are
left ``undecided''.  This notion introduced by
Chudnovsky~\cite{thesis,trigraphs} helps a lot to handle inductions,
especially when several kinds of decompositions appear in an arbitrary
order.

It is well known that an $O(n^k)$ algorithm that computes a maximum
weighted stable set for a class of perfect graphs closed under
complementation, yields an $O(n^{k+2})$ algorithm that computes an
optimal coloring.  See for instance~\cite{KrSe:colorP},
\cite{schrijver:opticomb} or Section~\ref{sec:color} below.  This
method, due to Gr\"ostchel, Lov\'asz, and Schrijver, is quite
effective and combinatorial.  Hence, from here on we just focus on an
algorithm that computes a maximum weighted stable set.  Also, in what
follows, in order to keep the paper as readable as possible, we
construct an algorithm that computes the weight of a maximum weighted
stable set, but does not output a set.  However, all our methods are
clearly constructive, so our algorithm may easily be turned into an
algorithm that actually computes the desired stable set.

Our algorithm may easily be turned into a \emph{robust} algorithm,
that is an algorithm that takes any graph as an input, and outputs
either a stable set on $k$ vertices and a partition of $V(G)$ into $k$
cliques of $G$ (so, a coloring of the complement), or some polynomial
size certificate proving that $G$ is not
``Berge-with-no-balanced-skew-partition''.  In the first case, we know
by the duality principle that the stable set is a maximum one, and the
clique cover is an optimal one, even if the input graph is not in the
class ``Berge-with-no-balanced-skew-partition''.  This feature is
interesting, because our algorithm is faster than the fastest one (so
far) for recognizing Berge graphs (Berge graphs can be recognized in
${O}(n^9)$ time~\cite{chudnovsky.c.l.s.v:reco}, and determining
whether a Berge graph has a balanced skew partition can be done in
${O}(n^5)$ time~\cite{nicolas:bsp,ChHaTrVu:2-join}).  However, it should be
pointed out that the certificate is not just and odd hole or antihole,
or a balanced skew-partition.  If the algorithm fails to find a stable
set at some point, the certificate is a decomposition tree, one leaf
of which satisfies none of the outputs of the decomposition theorem
for Berge graphs with no balanced skew-partitions; and if the
algorithm fails to find a clique cover, the certificate is a matrix on
$n+1$ rows showing that the graph is not perfect (see
Section~\ref{sec:color}).  This really certifies that a graph is not
in our class, but is maybe not as desirable as a hole, antihole, or
balanced skew-partition.

In Section~\ref{sec:def}, we give all the definitions and state some
known results. In Section~\ref{sec:blocks}, we define a new class of
Berge trigraphs called $\cal F$, and we prove a decomposition theorems
for trigraphs from $\cal F$.  In Section~\ref{sec:BlockDec}, we define
\emph{blocks} of decomposition. In Section~\ref{sec:bas}, we
show how to recognize all basic trigraphs, and find maximum weighted
stable sets for them. In
Section~\ref{sec:decAlpha}, we describe blocks of decomposition that
allow us to compute the maximum weight of a weighted stable set.  In
Section~\ref{sec:computeAlpha}, we give the main algorithm for
computing the maximum weight of a stable set in time $O(n^5)$.  

Results in the next sections are not needed to prove our main result.
We include them because they are of indepedent interest (while on the
same subject).  In Section~\ref{sec:color}, we describe the classical
algorithm that colors a perfect graph with a stable set oracle.  We
include it because it is hard to extract it from the deeper material that
surrounds it in \cite{gls:color} or~\cite{KrSe:colorP}.  In
Section~\ref{sec:ext}, we show that Berge trigraphs with no
balanced skew partitions admit extreme decompositions, that are
decompositions one block of decomposition of which is a basic
trigraph.  We do not need them here, but extreme decompositions are
sometimes very useful, in particular to prove properties by induction.
In Section~\ref{sec:end} we give an algorithm for finding extreme decompositions in a
trigraph (if any).  In Section~\ref{sec:enlarge}, we state several
open questions about how this work could be generalized to larger
classes of graphs.

We now state our main result (the formal definitions are given in the
next section, and the proof at the end of Section~\ref{sec:color}). In
complexity of algorithms, $n$ stands for the number of the vertices of
the input graph.

\begin{theorem}
  \label{th:colorM}
  There is an $O(n^7)$ time algorithm that colors any Berge graph with
  no balanced skew-partition.
\end{theorem}

\section{Trigraphs}
\label{sec:def}

For a set $X$, we denote by $X \choose 2$ the set of all subsets of
$X$ of size~2. For brevity of notation an element $\{ u,v \}$ of $X
\choose 2$ is also denoted by $uv$ or $vu$. A {\em trigraph} $T$
consists of a finite set $V(T)$, called the {\em vertex set} of $T$,
and a map $\theta : {{V(T)} \choose 2} \longrightarrow \{ -1,0,1 \}$,
called the {\em adjacency function}.

Two distinct vertices of $T$ are said to be {\em strongly adjacent} if
$\theta(uv)=1$, {\em strongly antiadjacent} if $\theta(uv)=-1$, and
{\em semiadjacent} if $\theta(uv)=0$. We say that $u$ and $v$ are
{\em adjacent} if they are either strongly adjacent, or semiadjacent;
and {\em antiadjacent} if they are either strongly antiadjacent, or
semiadjacent. An \emph{edge} (\emph{antiedge}) is a pair of adjacent
(antiadjacent) vertices. If $u$ and $v$ are adjacent (antiadjacent),
we also say that $u$ is {\em adjacent (antiadjacent) to} $v$, or that
$u$ is a {\em neighbor (antineighbor)} of $v$. Similarly, if $u$ and
$v$ are strongly adjacent (strongly antiadjacent), then $u$ is a {\em
  strong neighbor (strong antineighbor)} of $v$. Let $\eta(T)$ be the
set of all strongly adjacent pairs of $T$, $\nu(T)$ the set of all
strongly antiadjacent pairs of $T$, and $\sigma(T)$ the set of all
semiadjacent pairs of $T$. Thus, a trigraph $T$ is a graph if
$\sigma(T)$ is empty. A pair $\{u, v\} \subseteq V(T)$ of distinct
vertices is a \emph{switchable pair} if $\theta(uv) = 0$, a
\emph{strong edge} if $\theta(uv) = 1$ and a \emph{strong antiedge} if
$\theta(uv) = -1$.  An edge $uv$ (antiedge, strong edge, strong
antiedge, switchable pair) is \emph{between} two sets $A \subseteq
V(T)$ and $B \subseteq V(T)$ if $u\in A$ and $v \in B$ or if $u \in B$
and $v \in A$.

Let $T$ be a trigraph. The \emph{complement} $\overline{T}$ of $T$ is a
trigraph with the same vertex set as $T$, and adjacency function
$\overline{\theta}=-\theta$. For $v \in V(T)$, let $N(v)$ denote the
set of all vertices in $V(T) \setminus \{v\}$ that are adjacent to
$v$. Let $A \subset V(T)$ and $b
\in V(T) \setminus A$. We say that $b$ is {\em strongly complete} to
$A$ if $b$ is strongly adjacent to every vertex of $A$; $b$ is {\em
 strongly anticomplete} to $A$ if $b$ is strongly antiadjacent to
every vertex of $A$; $b$ is {\em complete} to $A$ if $b$ is adjacent
to every vertex of $A$; and $b$ is {\em anticomplete} to $A$ if $b$ is
antiadjacent to every vertex of $A$. For two disjoint subsets $A,B$
of $V(T)$, $B$ is {\em strongly complete (strongly anticomplete,
 complete, anticomplete)} to $A$ if every vertex of $B$ is strongly
complete (strongly anticomplete, complete, anticomplete) to $A$. A
set of vertices $X\subseteq V(T)$ \emph{dominates (strongly
 dominates)} $T$ if for all $v\in V(T)\setminus X$, there exists
$u\in X$ such that $v$ is adjacent (strongly adjacent) to $u$.

A {\em clique} in $T$ is a set of vertices all pairwise adjacent, and
a {\em strong clique} is a set of vertices all pairwise strongly
adjacent. A {\em stable set} is a set of vertices all pairwise
antiadjacent, and a {\em strongly stable set} is a set of vertices all
pairwise strongly antiadjacent. For $X \subset V(T)$ the trigraph
{\em induced by $T$ on $X$} (denoted by $T|X$) has vertex set $X$,
and adjacency function that is the restriction of $\theta$ to $X
\choose 2$. Isomorphism between trigraphs is defined in the natural
way, and for two trigraphs $T$ and $H$ we say that $H$ is an {\em
 induced subtrigraph} of $T$ (or $T$ {\em contains $H$ as an induced
 subtrigraph}) if $H$ is isomorphic to $T|X$ for some $X \subseteq
V(T)$. Since in this paper we are only concerned with the induced subtrigraph
containment relation, we say that \emph{$T$ contains~$H$} if $T$
contains $H$ as an induced subtrigraph. We denote by $T\setminus X$
the trigraph $T|(V(T) \setminus X)$.

Let $T$ be a trigraph. A \emph{path} $P$ of $T$ is a sequence of
distinct vertices $p_1, \dots, p_k$ such that either $k=1$, or for $i,
j \in \{1, \ldots, k\}$, $p_i$ is adjacent to $p_j$ if $|i-j|=1$ and
$p_i$ is antiadjacent to $p_j$ if $|i-j|>1$. Under these
circumstances, $V(P) = \{p_1, \dots, p_k\}$ and we say that $P$ is a
path {\em from $p_1$ to $p_k$}, its {\em interior} is the set
$P^*=V(P) \setminus \{p_1,p_k\}$, and the {\em length} of $P$ is
$k-1$. We also say that $P$ is a \emph{$(k-1)$-edge-path}. Sometimes,
we denote $P$ by $p_1 \c p_k$.  Observe that, since a graph is also a
trigraph, it follows that a path in a graph, the way we have defined
it, is what is sometimes in literature called a chordless path.

A {\em hole} in a trigraph $T$ is an induced subtrigraph $H$ of $T$
with vertices $h_1, \ldots, h_k $ such that $k \geq 4$, and for $i,j
\in \{1, \ldots, k\}$, $h_i$ is adjacent to $h_j$ if $|i-j|=1$ or
$|i-j|=k-1$; and $h_i$ is antiadjacent to $h_j$ if $1<|i-j|<k-1$. The
{\em length} of a hole is the number of vertices in it. Sometimes we
denote $H$ by $h_1 \c h_k \d h_1$. An {\em antipath} ({\em antihole})
in $T$ is an induced subtrigraph of $T$ whose complement is a path
(hole) in $\overline{T}$.

A {\em semirealization} of a trigraph $T$ is any trigraph $T'$ with
vertex set $V(T)$ that satisfies the following: for all $uv \in
{{V(T)} \choose 2}$, if $uv \in \eta(T)$ then $uv \in \eta(T')$, and
if $uv \in \nu(T)$ then $uv \in \nu(T')$.  Sometimes we will describe
a semirealization of $T$ as an {\em assignment of values} to
switchable pairs of $T$, with three possible values: ``strong edge'',
``strong antiedge'' and ``switchable pair''.  A {\em realization} of
$T$ is any graph that is semirealization of $T$ (so, any
semirealization where all switchable pairs are assigned the value
``strong edge'' or ``strong antiedge'').  For $S \subseteq \sigma
(T)$, we denote by $G^T_S$ the realization of $T$ with edge set $\eta
(T) \cup S$,  so in $G_{S}^T$ the switchable pairs in $S$ are assigned
the value ``edge'', and those in $\sigma(T) \setminus S$ the value
``antiedge''. The realization $G^T_{\sigma(T)}$ is called the {\em
  full realization} of~$T$.

Let $T$ be a trigraph. For $X \subseteq V(T)$, we say that $X$ and
$T|X$ are {\em connected} ({\em anticonnected}) if the graph
$G^{T|X}_{\sigma (T|X)}$ ($\overline{G^{T|X}_{\emptyset}}$) is
connected. A {\em connected component} (or simply \emph{component}) of
$X$ is a maximal connected subset of $X$, and an {\em anticonnected
component} (or simply \emph{anticomponent}) of $X$ is a maximal
anticonnected subset of $X$.

A trigraph $T$ is {\em Berge} if it contains no odd hole and no odd
antihole. Therefore, a trigraph is Berge if and only if its complement
is. We observe that $T$ is Berge if and only if every realization
(semirealization) of $T$ is Berge.

\subsection{Basic trigraphs}

A trigraph $T$ is {\em bipartite} if its vertex set can be partitioned 
into two strongly stable sets. Every realization of a bipartite 
trigraph is a bipartite graph, and hence every bipartite trigraph is Berge, 
and so is the complement of a bipartite trigraph.

A trigraph $T$ is a {\em line trigraph} if the full realization of $T$
is the line graph of a bipartite graph and every clique of size at
least $3$ in $T$ is a strong clique. The following is an easy fact
about line trigraphs.

\begin{theorem}
\label{linetrig}
If $T$ is a line trigraph, then every realization of $T$ is a line
graph of a bipartite graph.  Moreover, every semirealization of $T$ is
a line trigraph.
\end{theorem}

\Proof From the definition, the full realization $G$ of $T$ is a line
graph of a bipartite graph $R$. Let $S \subseteq \sigma(T)$. Define
$R_S$ as follows. For every $xy \in \sigma(T)\setminus S$, let
$v_{xy}$ be the common end of $x$ and $y$ in $R$. Then $v_{xy}$ has
degree 2 in $R$ because every clique of size at least $3$ in $T$ is a
strong clique. Let $a_{xy}$ and $b_{xy}$ be its neighbors. Now remove
$v_{xy}$ from $R$, and replace it by two new vertices, $u_{xy}$,
$w_{xy}$ such that $u_{xy}$ is only adjacent to $a_{xy}$, and $w_{xy}$
to $b_{xy}$. Then $R_S$ is bipartite and $G_S^T$ is the line graph of
$R_S$.  Hence, the first statement holds and the second follows
(because the full realization of a semirealization is a realization).
\bbox

Note that this implies that every line trigraph is Berge and so is the
complement of a line trigraph. Let us now define the trigraph analogue
of the double split graph (first defined in~\cite{CRST}), namely the
{\em doubled trigraph}.  A \emph{good partition} of a trigraph $T$ is
a partition $(X, Y)$ of $V(T)$ (possibly, $X=\emptyset$ or
$Y=\emptyset$) such that:

\begin{itemize}
\item Every  component of $T|X$ has at most two vertices, and every
  anticomponent of $T|Y$ has at most two vertices.
\item No switchable pair of $T$ meets both $X$ and $Y$. 
\item For every component $C_X$ of $T|X$, every anticomponent $C_Y$ of
  $T|Y$, and every vertex $v$ in $C_X \cup C_Y$, there exists at most
  one strong edge and at most one strong antiedge between $C_X$ and
  $C_Y$ that is incident to $v$.
\end{itemize}

A trigraph is \emph{doubled} if it has a good partition.  Doubled
trigraphs could also be defined as induced subtrigraphs of double
split trigraphs (see~\cite{trigraphs} for a definition of double split
trigraphs, we do not need it here).  Note that doubled trigraphs are
closed under taking induced subtrigraphs and complements (because $(X,
Y)$ is a good partition of some trigraph $T$ if and only if $(Y, X)$
is a good partition of $\overline{T}$).  A \emph{doubled graph} is any
realization of a doubled trigraph. We now show that:

\begin{theorem}
\label{bicotri} 
If $T$ is a doubled trigraph, then every realization of $T$ is a
doubled graph.  Moreover, every semirealization of $T$ is a doubled
trigraph.
\end{theorem}

\Proof The statement about realizations is clear from the
definition. Let $T$ be a doubled trigraph, and $(X, Y)$ a good
partition of $T$.  Let $T'$ be a semirealization of $T$.  It is easy
to see that $(X, Y)$ is also a good partition for $T'$ (for instance,
if a switchable pair $ab$ of $T|X$ is assigned value ``antiedge'',
then $\{a\}$ and $\{b\}$ become components of $T'|X$, but they still
satisfy the requirement in the definition of a good partition).
This proves the statement about semirealizations.  \bbox

Note that this implies that every doubled trigraph is Berge, because
every doubled graph is Berge.  Note that doubled graphs could be
defined equivalently as induced subgraphs of \emph{double split
  graphs} (see \cite{CRST} for a definition of double split graphs, we
do not need the definition here).

A trigraph is \emph{basic} if it is either a bipartite trigraph, the
complement of a bipartite trigraph, a line trigraph, the complement of
a line trigraph or a doubled trigraph.  The following sums up the
results of this subsection.

\begin{lemma}
  \label{l:presBas}
  Basic trigraphs are Berge, and are closed under taking induced
  subtrigraphs, semirealizations, realizations and complementation.
\end{lemma}

\subsection{Decompositions}

We now describe the decompositions that we need to state the
decomposition theorem. First, a {\em $2$-join} in a trigraph $T$ is a
partition $(X_1, X_2)$ of $V(T)$ such that there exist disjoint sets
$A_1, B_1, C_1, A_2, B_2, C_2 \subseteq V(T)$ satisfying:

\begin{itemize}
\item $X_1=A_1\cup B_1\cup C_1$ and $X_2=A_2\cup B_2\cup C_2$;
\item $A_1, A_2, B_1$ and $B_2$ are non-empty;
\item no switchable pair meets both $X_1$ and $X_2$;
\item every vertex of $A_1$ is strongly adjacent to every vertex of
  $A_2$, and every vertex of $B_1$ is strongly adjacent to every
  vertex of $B_2$;
\item there are no other strong edges between $X_1$ and $X_2$; 
\item for $i=1,2$ $|X_i| \geq 3$; and 
\item for $i = 1,2$, if $|A_i| = |B_i| = 1$, then the full realization of
$T|X_i$ is not a path of length two joining the members of $A_i$ and $B_i$.
\end{itemize}

In these circumstances, we say that $(A_1, B_1, C_1, A_2, B_2, C_2)$
is a \emph{split} of $(X_1, X_2)$. The $2$-join is \emph{proper} if
for $i = 1,2$, every component of $T|X_i$ meets both $A_i$ and
$B_i$. Note that the fact that a $2$-join is proper does not depend on
the particular split that is chosen. A \emph{complement $2$-join} of a
trigraph $T$ is a $2$-join of $\overline{T}$.  More specifically, a
{\em complement $2$-join} of a trigraph $T$ is a partition $(X_1, X_2)$
of $V(T)$ such that $(X_1, X_2)$ is a $2$-join of $\overline{T}$; and
$(A_1,B_1,C_1,A_2,B_2,C_2)$ is a {\em split} of this complement $2$-join
if it is a split of the respective $2$-join in the complement, i.e.\
$A_1$ is strongly complete to $B_2\cup C_2$ and strongly anticomplete
to $A_2$, $C_1$ is strongly complete to $X_2$, and $B_1$ is strongly
complete to $A_2 \cup C_2$ and strongly anticomplete to $B_2$.

\begin{theorem}
 \label{l:par2Join} Let $T$ be a Berge trigraph and $(A_1, B_1, C_1,
A_2, B_2, C_2)$ a split of a proper $2$-join of $T$. Then all paths
with one end in $A_i$, one end in $B_i$ and interior in $C_i$, for
$i=1, 2$, have lengths of the same parity.
\end{theorem}

\Proof Otherwise, for $i=1, 2$, let $P_i$ be a path with one end in
$A_i$, one end in $B_i$ and interior in $C_i$, such that $P_1$ and
$P_2$ have lengths of different parity. They form an odd hole, a
contradiction. \bbox

Our second decomposition is the balanced skew-partition. Let $A,B$
be disjoint subsets of $V(T)$. We say the pair $(A,B)$ is {\em
 balanced} if there is no odd path of length greater than $1$ with
ends in $B$ and interior in $A$, and there is no odd antipath of
length greater than $1$ with ends in $A$ and interior in $B$. A {\em
 skew-partition} is a partition $(A,B)$ of $V(T)$ so that $A$ is not
connected and $B$ is not anticonnected. A skew-partition $(A,B)$ is {\em balanced} if the pair $(A,B)$ is. 
Given a balanced skew-partition $(A,B)$, $(A_1,A_2,B_1,B_2)$ is a
\emph{split of $(A,B)$} if $A_1, A_2, B_1$ and $B_2$ are disjoint
non-empty sets, $A_1 \cup A_2 =A$, $B_1 \cup B_2=B$, $A_1$ is strongly
anticomplete to $A_2$, and $B_1$ is strongly complete to $B_2$. Note
that for every balanced skew-partition, there exists at least one
split.

The two decompositions we just described generalize some
decompositions used in~\cite{CRST}, and in addition all the
``important'' edges and non-edges in those graph decompositions are
required to be strong edges and strong antiedges of the trigraph,
respectively.  We now state several technical lemmas.

A trigraph is called {\em monogamous} if every vertex of it belongs to
at most one switchable pair. We are now ready to state the
decomposition theorem for Berge monogamous trigraphs. This is
Theorem~3.1 of \cite{trigraphs}.
\begin{theorem} 
\label{noMjointri}
Let $T$ be a monogamous Berge trigraph. Then one of the following holds:
\begin{itemize}
\item $T$ is basic;
\item $T$ or $\overline{T}$ admits a proper $2$-join; or 
\item $T$ admits a balanced skew-partition.
\end{itemize}
\end{theorem}

When $(A, B)$ is a skew-partition of a trigraph $T$, we say that $B$ is
a \emph{star cutset} of $T$ if at least one anticomponent of $B$ has size~1.
The following is Theorem~5.9 from~\cite{thesis}.

\begin{theorem}
 \label{starcutset}
 If a Berge trigraph admits a star cutset, then it admits a balanced
 skew-partition.
\end{theorem}

Let us say that $X$ is a \emph{homogeneous set} in
a trigraph $T$ if $1<|X|<|V(T)|$, and every vertex of $V(T) \setminus X$ is
either strongly complete or strongly anticomplete to $X$.

\begin{theorem}
\label{lemma}
Let $T$ be a trigraph and let $X$ be a homogeneous set in $T$, 
such that some vertex of $V(T) \setminus X$ is strongly complete to $X$, and
some vertex of $V(T) \setminus X$ is strongly anticomplete to $X$. Then
$T$ admits a balanced skew-partition.
\end{theorem}

\Proof Let $A$ be the set of vertices of $V(T) \setminus X$ that are 
strongly anticomplete to $X$, and $C$ the set of vertices of 
$V(T) \setminus X$ that are strongly complete to $X$. Let $x \in X$. Then 
$C \cup \{x\}$ is a star cutest of $T$ (since $A$ and $X \setminus \{x\}$ 
are non-empty and strongly anticomplete to each other), and so $T$ 
admits a balanced skew-partition by~\ref{starcutset}. \bbox

We also need the following (this is an immediate corollary of Theorem 5.13 in 
\cite{thesis}):
\begin{theorem}
\label{onepair}
Let $T$ be a Berge trigraph.
Suppose that there is a partition of $V(T)$ into four nonempty sets 
$X,Y,L,R$, such that $L$ is strongly anticomplete to $R$, and $X$
is strongly complete to $Y$. If $(L,Y)$ is balanced then $T$ admits a balanced 
skew-partition.
\end{theorem}

\section{Decomposing trigraphs from $\mathcal F$}
\label{sec:blocks}

Let $T$ be a trigraph, denote by $\Sigma(T)$ the graph with vertex set
$V(T)$ and edge set $\sigma(T)$ (the switchable pairs of $T$).  The
connected components of $\Sigma(T)$ are called the \emph{switchable
  components} of $T$.  Let $\mathcal{F}$ be the class of Berge
trigraphs such that the following hold:
\begin{itemize}
\item Every switchable component of $T$ has at most two edges (and
therefore no vertex has more than two neighbors in $\Sigma(T)$).
\item Let $v \in V(T)$ have degree two in $\Sigma(T)$, denote its
neighbors by $x$ and~$y$. Then either $v$ is strongly complete to
$V(T) \setminus \{v, x, y\}$ in $T$, and $x$ is strongly adjacent to
$y$ in $T$ (in this case we say that $v$ and the switchable component
that contains $v$ are {\em heavy}), or $v$ is
strongly anticomplete to $V(T) \setminus \{v, x, y\}$ in $T$, and $x$
is strongly antiadjacent to $y$ in $T$ (in this case we say that $v$
and the switchable component that contains $v$ are {\em light}).
\end{itemize} 

Observe that $T\in \mathcal{F}$ if and only if $\overline{T}\in
\mathcal{F}$; also $v$ is light in $T$ if and only if $v$ is heavy in
$\overline{T}$.

\begin{theorem}\label{2joinform}
  Let $T$ be a trigraph from $\mathcal F$ with no balanced
  skew-partition, and let $(A_1,B_1,C_1,A_2,B_2,C_2)$ be a split of a
  $2$-join $(X_1,X_2)$ in $T$. Then the following hold:
 \begin{enumerate}[(i)]
 \item $(X_1,X_2)$ is a proper $2$-join;
 \item every vertex of $X_i$ has a neighbor in $X_i$, $i=1,2$;
 \item every vertex of $A_i$ has an antineighbor in $B_i$, $i=1,2$;
 \item every vertex of $B_i$ has an antineighbor in $A_i$, $i=1,2$;
 \item every vertex of $A_i$ has a neighbor in $C_i\cup B_i$,
  $i=1,2$;
 \item every vertex of $B_i$ has a neighbor in $C_i\cup A_i$,
  $i=1,2$;
\item if $C_i = \emptyset$, then $|A_i| \geq 2$ and $|B_i| \geq 2$,
  $i=1,2$;
\item \label{size}$|X_i| \geq 4$, $i=1,2$.
 \end{enumerate}
\end{theorem}

\Proof 
Note that by \ref{starcutset}, neither $T$ nor $\overline{T}$ can have
a star cutset.  

To prove $(i)$, we just have to prove that every component of $T|X_i$
meets both $A_i$ and $B_i$, $i=1,2$. Suppose for a contradiction that
some connected component $C$ of $T|X_1$ does not meet $B_1$ (the other
cases are symmetric). If there is a vertex $c\in C\setminus A_1$ then
for any vertex $u\in A_2$, we have that $\{u\}\cup A_1$ is a star
cutset that separates $c$ from $B_1$, a contradiction.  So,
$C\subseteq A_1$. If $|A_1|\geq 2$ then pick any vertex $c\in C$ and
$c'\not= c$ in $A_1$. Then $\{c'\}\cup A_2$ is a star cutset that
separates $c$ from $B_1$. So, $C=A_1=\{c\}$. Hence, there exists some
component of $T|X_1$ that does not meet $A_1$, so by the same argument
as above we deduce $|B_1|=1$ and the unique vertex of $B_1$ has no
neighbor in $X_1$. Since $|X_1|\geq 3$, there is a vertex $u$ in
$C_1$. Now $\{c ,a_2\}$ where $a_2\in A_2$ is a star cutset that
separates $u$ from $B_1$, a contradiction.

To prove $(ii)$, just notice that if some vertex in $X_i$ has no
neighbor in $X_i$, then it forms a component of $T|X_i$ that does not
meet one of $A_i, B_i$. This is a contradiction to $(i)$.

To prove $(iii)$ and $(iv)$, consider a vertex $a\in A_1$ strongly
complete to $B_1$ (the other cases are symmetric). If $A_1\cup
C_1\not= \{a\}$ then $B_1\cup A_2\cup\{a\}$ is a star cutset that
separates $(A_1\cup C_1)\setminus \{a\}$ from $B_2$. So $A_1\cup
C_1=\{a\}$ and $|B_1|\geq 2$ because $|X_1|\geq 3$.  But now $B_1$ is
a homogeneous set, strongly complete to $A_1$ and strongly
anticomplete to $A_2$, and so $T$ admits a balanced skew-partition
by~\ref{lemma}, a contradiction.

To prove $(v)$ and $(vi)$, consider a vertex $a\in A_1$ strongly
anticomplete to $C_1\cup B_1$ (the other cases are symmetric). By
$(ii)$, $a$ has a neighbor in $A_1$, and so $A_1\not = \{a\}$. But now
$\{a\}\cup B_1\cup C_1\cup B_2\cup C_2$ is a star cutset in
$\overline{T}$, a contradiction.

To prove $(vii)$, suppose that $C_1=\emptyset$ and $|A_1|=1$ (the
other cases are symmetric). By $(iv)$ and $(vi)$, and since
$C_1=\emptyset$, $A_1$ is both complete and anticomplete to
$B_1$. This implies that the unique vertex of $A_1$ is semiadjacent
to every vertex of $B_1$, and therefore, since $T\in\mathcal{F},
|B_1|\leq 2$. Since $|X_1|\geq 3$, we deduce that $|B_1|=2$, and,
since $T\in \mathcal{F}$, the unique vertex of $A_1$ is either
strongly complete or strongly anticomplete to $V(T)\setminus(A_1\cup
B_1)$, which is a contradiction because $A_1$ is strongly complete to
$A_2$ and strongly anticomplete to $B_2$.

To prove $(viii)$, we may assume by $(vii)$ that $C_1 \neq \emptyset$,
so suppose for a contradiction that $|A_1|=|C_1|=|B_1|=1$.  Let $a, b,
c$ be the vertices in $A_1, B_1, C_1$ respectively.  By $(iii)$, $ab$
is an antiedge.  Also, $c$ is adjacent to $a$, for otherwise, there is
a star cutset centered at $b$ that separates $a$ from $c$. Similarly,
$c$ is adjacent to $b$.  Since the full realization of $T|X_1$ is not
a path of length 2 from $a$ to $b$, we know that $ab$ is a switchable
pair.  But this contradicts~\ref{l:par2Join}.
\bbox

Let $b$ be a vertex of degree two in $\Sigma(T)$, and let $a,c$ be the
neighbors of $b$ in $\Sigma(T)$. Assume that $b$ is light. We call a
vertex $w \in V(T) \setminus \{a, b, c\}$ an {\em $a$-appendage} of
$b$ if there exist $u,v \in V(T) \setminus \{a,b,c\}$ such that:
\begin{itemize}
\item $a \d u \d v \d w$ is a path;
\item $u$ is strongly anticomplete to $V(T) \setminus \{a,v\}$;
\item $v$ is strongly anticomplete to $V(T) \setminus \{u,w\}$; and
\item $w$ has no neighbors in $\Sigma(T)$ except possibly $v$ (i.e.\
  there is no switchable pair containing $w$ in $T$ except possibly
  $vw$).
\end{itemize}

A $c$-appendage is defined similarly. If $b$ is a heavy vertex of $T$,
then $w$ is an $a$-appendage of $b$ in $T$ if and only if $w$ is an
$a$-appendage of $b$ in $\overline{T}$.

The following is an analogue of~\ref{noMjointri} for trigraphs in $\mathcal{F}$.  It can 
be easily deduced from~\cite{thesis}, but for the reader's convenience we include a
short proof, whose departure point is~\ref{noMjointri}.

\begin{theorem}
  \label{structure}
  Every trigraph in $\mathcal{F}$ is either basic, or admits a
  balanced skew-partition, a proper $2$-join, or a proper $2$-join
  in the complement.
\end{theorem}

\Proof For $T \in \mathcal{F}$, let $\tau(T)$ be the number of
vertices of degree two in $\Sigma(T)$. The proof is by induction on
$\tau(T)$. If $\tau(T)=0$, then the result follows
from~\ref{noMjointri}. Now let $T \in \mathcal{F}$ and let $b$ be a
vertex of degree two in $\Sigma(T)$. Let $a,c$ be the two neighbors of
$b$ in $\Sigma(T)$. By passing to the complement if necessary, we may
assume that $b$ is light.

Let $T'$ be the trigraph obtained from $T$ by making $a$ strongly
adjacent to $b$. If $b$ has no $a$-appendages, then no further changes
are necessary; set $W=\emptyset$. Otherwise choose an $a$-appendage
$w$ of $b$, and let $u,v$ be as in the definition of an $a$-appendage;
set $V(T')=V(T) \setminus \{u,v\}$ and make $a$ semiadjacent to $w$
in $T'$; set $W=\{w\}$.

If $W=\emptyset$ then clearly $T'\in {\cal F}$ and $\tau (T)> \tau
(T')$. Suppose that $W\neq \emptyset$. If $t \in V(T')$ is adjacent to
both $a$ and $w$, then $a \d u \d v \d w \d t$ is an odd hole in
$T$. Thus no vertex of $T'$ is adjacent to both $a$ and $w$. In
particular, no antihole of length at least $7$ of $T'$ goes through
$a$ and $w$.  Also, there is no odd hole that goes through $a$ and
$w$.  Hence $T'$ is in $\mathcal{F}$. Moreover, $\tau(T)>\tau(T')$ (we
remind the reader that $v$ is the only possible neighbor of $w$ in
$\Sigma(T)$).

Inductively, one of the outcomes of~\ref{structure} holds for $T'$.
We consider the following cases, and show that in each of them, one of
the outcomes of~\ref{structure} holds for $T$.

\noindent{\bf Case 1:} $T'$ is basic.

Suppose first that $T'$ is bipartite.  We claim that $T$ is bipartite.
Let $V(T')=X \cup Y$ where $X$ and $Y$ are disjoint strongly stable
sets. The claim is clear if $b$ has no $a$-appendage, so we may
assume that $W=\{w\}$. We may assume that $a \in X$; then $w \in
Y$. Then $X \cup \{v\}$ and $Y \cup \{u\}$ are strongly stable sets of
$T$ with union $V(T)$, and thus $T$ is bipartite.

Suppose $T'$ is a line trigraph.  First observe that no clique of size
at least three in $T$ contains $u,v$ or $b$. So, if $W=\emptyset$,
then clearly $T$ is a line trigraph. So assume that $W\neq
\emptyset$. Note that the full realization of $T$ is obtained from the
full realization of $T'$ by subdividing twice the edge $aw$. Since no
vertex of $T'$ is adjacent to both $a$ and $w$, it follows that $T$ is a
line trigraph (because line graphs are closed under subdividing an
edge whose ends have no common neighbors, and line graphs of bipartite
graphs are closed under subdividing twice such an edge).

Suppose $\overline{T'}$ is bipartite, and let $X,Y$ be a partition of
$V(T)$ into two strong cliques of $T'$. We may assume that $a \in
X$. Assume first that $b \in Y$. Since $a$ is the unique strong
neighbor of $b$ in $T'$, it follows that $Y=\{b\}$, so $X$ contains
$a$ and $c$, a contradiction. Thus we may assume that $b \in X$. Since
$a$ is the unique strong neighbor of $b$ in $T'$, it follows that
$X=\{a,b\}$, and $b$ is strongly anticomplete to $Y \setminus
\{c\}$. Let $N$ be the set of strong neighbors of $a$ in $Y \setminus
\{c\}$, and $M$ the set of strong antineighbors of $a$ in $Y \setminus
\{c\}$. Since $T \in \mathcal{F}$, it follows that $Y=N \cup M \cup W
\cup \{c\}$. If either $|N|>1$ or $|M|>1$, then $T$ admits a balanced
skew-partition by~\ref{lemma}, so we may assume that $|N| \leq 1$ and
$|M| \leq 1$. Since no vertex of $T'$ is adjacent to both $a$ and $w$,
it follows that $|N \cup W| \leq 1$. Now if $M=\emptyset$ or $N\cup
W=\emptyset$ then $T'$ is bipartite and we proceed as above, otherwise
$N\cup W\cup \{c\}$ is a clique cutset of $T'$ of size $2$, which is a
star cutset in $T$, and hence $T$ admits a balanced skew-partition
by~\ref{starcutset}.

Next assume that $\overline{T'}$ is a line trigraph. Since $bc$ is a
switchable pair in $T'$ and $b$ is strongly anticomplete to $V(T')
\setminus \{a,b,c\}$, it follows that $c$ is strongly complete to
$V(T') \setminus \{a,b,c\}$ else there would be in $\overline{T'}$ a
clique of size $3$ with a switchable pair. Since $\overline{T'}$ is
a line trigraph, it follows that for every triangle $S$ of $T'$ and a
vertex $v \in V(T') \setminus S$, $v$ has at least one strong neighbor
in $S$. If $x,y \in V(T') \setminus \{a,b,c\}$ are adjacent, then
$\{x,y,c\}$ is a triangle and $b$ has no strong neighbor in it, and
hence $V(T') \setminus \{a,b,c\}$ is a strongly stable set. But now,
$V(T') \setminus \{a,c\}, \{a,c\}$ form a partition of $V(T')$ into
two strongly stable sets of $T'$.  So $T'$ is bipartite and we proceed as
above.

Finally, suppose that $T'$ is doubled and let $(X, Y)$ be a good
partition of~$T'$.  If $T'|Y$ is empty or has a unique anticomponent,
then $T'$ is bipartite.  Hence, we may assume that $Y$ contains two
strongly adjacent vertices $x$ and~$x'$.  If there exist $y\neq x$ and
$y'\neq x'$ such that $\{x, y\}$ and $\{x', y'\}$ are anticomponents
of $T'|Y$, then every vertex of $T'$ has at least two strong
neighbors, a contradiction because of $b$.  It follows that $\{x\}$,
say, is an anticomponent of $T'|Y$.  If $T'|X$ has a single component or
is empty, then $T'$ is the complement of a bipartite trigraph.  Hence
we may assume that $T'|X$ has at least two components.  Therefore, $Y$
is a star cutset of $T'$ centered at $x$.  This is handled in the next
case.

\noindent{\bf Case 2:} $T'$ admits a balanced skew-partition.

Let $(A,B)$ be a balanced skew-partition of $T'$. If $W \neq
\emptyset$, let $A'=A \cup \{u,v\}$; and if $W=\emptyset$, let $A'=A$.
Then $T|A'$ is not connected. We claim that if some anticomponent $Y$
of $B$ is disjoint from $\{a,b\}$, then $T$ admits a balanced skew-partition. Since $a$ is complete to $W$ in $T'$, some component $L$ of
$A$ is disjoint from $\{ a\} \cup W$, and hence $L$ is a component of
$A'$ as well.  We may assume w.l.o.g. that $Y$ is disjoint from $W$
(this is clearly the case if $B\cap \{ a,b \} \neq \emptyset$, and if
$B\cap \{ a,b \} = \emptyset$ we may assume w.l.o.g. that $Y\cap
W=\emptyset$).  Now, in $T$, $Y$ is strongly complete to $B \setminus
Y$, $L$ is strongly anticomplete to $A' \setminus L$, and thus
$(A',B)$ is a skew-partition of $T$ and $(L\cup Y) \cap (\{ a,b \}
\cup W\cup (A'\setminus A)) \subseteq \{ b \}$.  Since $(A,B)$ is a
balanced skew-partition of $T'$, the pair $(L,Y)$ is balanced in $T$;
consequently~\ref{onepair} implies that $T$ admits a balanced skew-partition. This proves the claim.

Thus we may assume that no such $Y$ exists, and therefore $T'|B$ has
exactly two anticomponents, $B_1$ and $B_2$, and $a \in B_1$ and $b
\in B_2$. Since $a$ is the unique strong neighbor of $b$ in $T'$, it
follows that $B_1=\{a\}$. Since $a$ is anticomplete to $W \cup \{c\}$,
we deduce that $W \cup \{c\} \subseteq A'$. Let $A_1$ be the component
of $T|A'$ containing $c$ and $A_2=A'\setminus A_1$. Suppose that $a$
does not have a strong neighbor in $T$.  Then $B_2=\{ b\}$, and since
$T\in {\cal F}$, $a$ is strongly anticomplete to $A'$. We may assume
that $T$ is not bipartite, since otherwise $T$ satisfies one of the
outcomes of the theorem we are proving. Then $T$ contains an odd cycle
$C$, which must be in $A_1$ or $A_2$ (since $\{ a,b\}$ is strongly
anticomplete to $A'$).  Since $T \in {\cal F}$, $C$ must contain at
least one strong edge, say $xy$. But then $\{ x,y \}$ is a star cutest
in $T$ separating $\{ a,b \}$ from a vertex of $A_2$. So
by~\ref{starcutset}, $T$ has a balanced skew-partition. Therefore we
may assume that $a$ has at least one strong neighbor in~$T$.

Let $x\in A_2$. Let $N$ be the set of strong neighbors of $a$ in
$T$. Then $(N \cup \{a\}) \setminus \{x\}$ is a star cutset in $T$
separating $b$ from $x$, unless $x$ is the unique strong neighbor of
$a$. In this case $\{a,x\}$ is a star cutset separating $A_1$ from
$A_2\setminus\{x\}$, unless $A_2=\{x\}$. Now suppose that $c$ has a
neighbor $y$ (that is in fact a strong neighbor since $T \in {\cal
  F}$). Then $\{c,y\}$ is a star cutset separating $A_1 \setminus \{c,
y\}$ from $x$, unless $A_1 = \{c, y\}$, in which case $T$ is
bipartite. So we may assume that $c$ has no neighbor in $A_1$.  Now,
either $T$ is bipartite, or $T$ has an odd cycle. But in this later
case, the cycle is in $A_1$ and any strong edge of it (which exists
since $T\in {\cal F}$) forms a star cutset separating $c$ from the
rest of the cycle.  Therefore, by~\ref{starcutset}, $T$ has a balanced
skew-partition.

\noindent{\bf Case 3:} $T'$ admits a proper $2$-join.

Let $(A_1,B_1,C_1,A_2,B_2,C_2)$ be a split of a proper $2$-join of
$T'$. We may assume that $a \in A_1 \cup B_1 \cup C_1$. Then $W
\subseteq A_1 \cup B_1 \cup C_1$. If $W \neq \emptyset$ let $C_1'=C_1
\cup \{u,v\}$, and otherwise let $C_1'=C_1$. We may assume that
$(A_1,B_1,C_1',A_2,B_2,C_2)$ is not a proper $2$-join of $T$, and
hence w.l.o.g.\ $a \in A_1$ and $b \in A_2$. Then $c \in B_2 \cup
C_2$. Since $a$ is the unique strong neighbor of $b$ in $T'$, it
follows that $A_1=\{a\}$. By Case~2, we may assume that $T'$ does not
admit a balanced skew-partition, and therefore~\ref{2joinform} implies
that $a$ is anticomplete to $B_1$.  Note that since $T \in {\cal F}$,
$ab$ is the only switchable pair in $T$ that involves $a$.  Let $N$ be
the set of strong neighbors of $a$ in $C_1'$ in $T$. It follows from
the definition of a proper $2$-join that $N \neq \emptyset$.  We may
assume that $T$ does not admit a balanced skew-partition, and hence
by~\ref{2joinform}, every $2$-join of $T$ is proper.  So either $(N,B_1,
C_1' \setminus N,\{a\}, B_2, C_2 \cup A_2)$ is a split of a proper
$2$-join in $T$, or $|N|=|B_1|=1$ and the full realization of $T|(C_1'
\cup B_1)$ is a path of length two joining the members of $N$ and
$B_1$. Let this path be $n \d n' \d b_1$ where $n \in N$ and $b_1 \in
B_1$. Since $b_1$ has no neighbor in $\Sigma(T)$ except possibly $n'$,
it follows that $b_1$ is an $a$-appendage of $b$. In particular, $W
\neq \emptyset$. Since $W \subseteq B_1 \cup C_1$, it follows that
$w=b_1$, $u=n$ and $v=n'$. But then $|A_1 \cup B_1 \cup C_1|=2$,
contrary to the fact that $(A_1,B_1,C_1,A_2,B_2,C_2)$ is a split of a
proper $2$-join of $T'$.

\noindent{\bf Case 4:} $\overline{(T')}$ admits a proper $2$-join.

Let $(A_1,B_1,C_1,A_2,B_2,C_2)$ be a split of a proper $2$-join in
$\overline{T'}$. First suppose that $W\neq \emptyset$. Then we may
assume that $a,w \in A_1 \cup B_1\cup C_1$. Since no vertex of $T'$ is
adjacent to both $a$ and $w$, it follows w.l.o.g. that $a\in A_1$, $w
\in B_1$ and $C_2=\emptyset$.  Since $a$ is the unique strong neighbor
of $b$ in $T'$, it follows that $b \in B_2$ and $C_1=\emptyset$.  But
now $(A_1,B_1,\emptyset , B_2, A_2, \emptyset)$ is a split of a
$2$-join in $T'$. By Case 2 we may assume that $T'$ does not admit a
balanced skew-partition, and hence this $2$-join is proper
by~\ref{2joinform}. But then we may proceed as in Case 3.  Therefore
we may assume that $W=\emptyset$.

We may assume that $(A_1,B_1,C_1, A_2,B_2, C_2)$ is
not a split of a proper $2$-join in $\overline{T}$, and therefore $a
\in A_1 \cup B_1 \cup C_1$, and $b \in A_2 \cup B_2 \cup C_2$ (up to
symmetry). Since $a$ is the unique strong neighbor of $b$ in $T'$, and
since $A_1,B_1$ are both non-empty, we deduce that $b \not \in C_2$,
and so we may assume that $b \in B_2$. Since $A_1 \neq \emptyset$, it
follows that $C_1 = \emptyset$ and $A_1=\{a\}$.  Since $|A_1 \cup B_1 \cup C_1|
\geq 3$, it follows that $|B_1|\geq 2$.  Since $c$ is strongly
antiadjacent to $a$ and semiadjacent to $b$ in $T$, we deduce that $c \in
A_2$. But now, if $a$ has a neighbor $x \in B_1$ in $T$ (which is therefore a
strong neighbor), then $\{x,a\} \cup A_2 \cup C_2$ is a star cutset
in $T$, and if $a$ is strongly anticomplete to $B_1$ in $T$, then it follows
from the definition of a proper $2$-join that $B_1$ is a 
homogeneous set in~$T$. In both cases, by~\ref{starcutset} and
\ref{lemma}, respectively, we deduce that $T$ admits a balanced
skew-partition.
\bbox

\section{Blocks of decomposition}
\label{sec:BlockDec}

The way we use decompositions for computing stable sets in
Section~\ref{sec:decAlpha} requires building blocks of decomposition
and asking several questions on the blocks. To do that
we need to ensure that the blocks of decomposition are still in our class.
 
A set $X\subseteq V(T)$ is a \emph{fragment} of a trigraph $T$ if
one of the following holds:
\begin{enumerate}
\item\label{i:2J} $(X,V(T)\setminus X)$ is a proper $2$-join of $T$;
\item\label{i:C2J} $(X,V(T)\setminus X)$ is a proper complement $2$-join of $T$.
\end{enumerate}

Note that a fragment of $T$ is a fragment of $\overline{T}$. We now
define the \emph{blocks of decomposition $T_X$} with respect to some
fragment $X$.  A $2$-join is \emph{odd} or \emph{even} according to
the parity of the lengths of the paths described in~\ref{l:par2Join}.

If $(X_1,X_2)$ is a proper odd $2$-join and $X=X_1$, then let
$(A_1,B_1,C_1,A_2,B_2,C_2)$ be a split of $(X_1, X_2)$. We build the
block of decomposition $T_{X_1}=T_X$ as follows. We start with $T |(
A_1\cup B_1\cup C_1)$. We then add two new \emph{marker vertices} $a$
and $b$ such that $a$ is strongly complete to $A_1$, $b$ is strongly
complete to $B_1$, $ab$ is a switchable pair, and there are no other 
edges between $\{a,b\}$ and $X_1$.  Note that $\{a, b\}$
is a switchable component of $T_X$.  We call it the \emph{marker
  component of~$T_X$}.   

If $(X_1,X_2)$ is a proper even $2$-join and $X=X_1$, then let
$(A_1,B_1,C_1,A_2,B_2,C_2)$ be a split of $(X_1, X_2)$. We build the
block of decomposition $T_{X_1}=T_X$ as follows. We start with $T |(
A_1\cup B_1\cup C_1)$. We then add three new \emph{marker vertices}
$a$, $b$ and $c$ such that $a$ is strongly complete to $A_1$, $b$ is
strongly complete to $B_1$, $ac$ and $cb$ are switchable pairs, and
there are no other edges between $\{a, b, c\}$ and $X_1$.  Again,
$\{a, b, c\}$ is called the \emph{marker component of~$T_X$}.

If $(X_1,X_2)$ is a proper odd complement $2$-join and $X=X_1$, then
let $(A_1,B_1,C_1,A_2,B_2,C_2)$ be a split of $(X_1, X_2)$. We build
the block of decomposition $T_{X_1}=T_X$ as follows. We start with $T
|( A_1\cup B_1\cup C_1)$. We then add two new \emph{marker vertices}
$a$ and $b$ such that $a$ is strongly complete to $B_1\cup C_1$, $b$
is strongly complete to $A_1\cup C_1$, $ab$ is a switchable pair, and
there are no other edges between $\{a, b\}$ and $X_1$.  Again, $\{a,
b\}$ is called the \emph{marker component of~$T_X$}.

If $(X_1,X_2)$ is a proper even complement $2$-join and $X=X_1$, then
let $(A_1,B_1,C_1,A_2,B_2,C_2)$ be a split of $(X_1, X_2)$. We build
the block of decomposition $T_{X_1}=T_X$ as follows. We start with $T
|( A_1\cup B_1\cup C_1)$. We then add three new \emph{marker vertices}
$a$, $b$ and $c$ such that $a$ is strongly complete to $B_1\cup C_1$,
$b$ is strongly complete to $A_1\cup C_1$, $c$ is strongly complete to
$X_1$, $ac$ and $cb$ are switchable pairs, $ab$ is a strong edge, and
there are no other edges between $\{a, b, c\}$ and $X_1$.  Again,
$\{a, b, c\}$ is called the \emph{marker component of~$T_X$}.

\begin{theorem}
 \label{l:stayBerge}
 If $X$ is a fragment of a trigraph $T$ from $\mathcal F$ with no
 balanced skew-partition, then $T_X$ is a trigraph from~$\mathcal F$.
\end{theorem}

\Proof From the definition of $T_X$, it is clear that every vertex of
$T_X$ is in at most one switchable pair, or is heavy, or is light.
So, to prove that $T_X \in {\mathcal F}$, it remains only to prove
that $T_X$ is Berge.

Let $X=X_1$ and $(X_1,X_2)$ is a proper $2$-join of $T$. Let
$(A_1,B_1,C_1,A_2,B_2,C_2)$ be a split of $(X_1,X_2)$. 

Suppose first that $T_{X_1}$ has an odd hole $H = h_1 \c h_k \d h_1$.
Assume that the vertices of $Z_{X_1}$ are consecutive in $H$,
then $H\setminus Z_{X_1}$ is a path $P$ with one end in $A_1$, the
other one in $B_1$ and interior in $C_1$. A hole of $T$ is obtained
by adding to $P$ a path with one end in $A_2$, the other one in $B_2$,
and interior in $C_2$. By~\ref{l:par2Join}, this hole is odd, a
contradiction. Thus the marker vertices are not consecutive in $H$,
and since $c$ has no neighbors in $V(T)\setminus\{a,b,c\}$, we deduce
that $c\not \in V(H)$. Now a hole of the same length as $H$ is obtained in
$T$ by possibly replacing $a$ and/or $b$ by some vertices $a_2 \in
A_2$ and $b_2\in B_2$, chosen to be antiadjacent (this is possible
by~\ref{2joinform}).

Suppose now that $T_{X_1}$ has an odd antihole $H = h_1 \c h_k \d
h_1$. Since an antihole of length~5 is also a hole, we may assume that
$H$ has length at least~7. So, in $H$, any pair of vertices has a
common neighbor. It follows that at most one of $a, b, c$ is in $H$,
and because of its degree, $c$ is not in $H$. 
An antihole of same length as $H$ is obtained in
$T$ by possibly replacing $a$ or $b$ by some vertices $a_2 \in
A_2$ or $b_2\in B_2$, a contradiction. 

Note that the case when $T$ has a complement $2$-join follows by complementation.

\bbox

\begin{theorem}\label{stab}
  If $X$ is a fragment of a trigraph $T$ from $\mathcal{F}$ with no
  balanced skew-partition, then the block of decomposition $T_X$ has
  no balanced skew-partition.
\end{theorem}

\Proof To prove this, we suppose that $T_X$ has a balanced skew-partition $(A',B')$ with a split $(A_1',A_2',B_1',B_2')$.  From this,
we find a skew-partition in $T$.  Then we use~\ref{onepair} to prove
the existence of a \emph{balanced} skew-partition in $T$.  This gives
a contradiction that proves the theorem.

Let $X=X_1$ and $(X_1,X_2)$ is a proper $2$-join of
$T$. Let $(A_1,B_1,C_1,A_2,B_2,C_2)$ be a split of $(X_1,X_2)$.

Since the marker vertices in $T_X$, $a$ and $b$ have no common strong
neighbor and $c$ has no strong neighbor, there are up to symmetry two
cases: $a\in A_1'$ and $b\in A_1'$, or $a\in A_1'$ and $b\in
B_1'$. Note that when $(X_1, X_2)$ is even, the marker vertex $c$ must
be in $A'_1$ because it is adjacent to $a$ and has no strong neighbor.

Assume first that $a$ and $b$ are both in $A_1'$. Then $(X_2 \cup A_1'
\setminus Z_{X_1}, A_2', B_1', B_2')$ is a split of a skew-partition $(A,B)$ in~$T$.  The pair $(A_2', B_1')$ is
balanced in $T$ because it is balanced in $T_X$.  Hence,
by~\ref{onepair}, $T$ admits a balanced skew-partition, a
contradiction.

Thus not both $a$ and $b$ are in $A_1'$, and so $a\in A_1'$ and $b\in
B_1'$. In this case, $(A_2 \cup C_2 \cup A_1' \setminus \{a, c\},
A_2', B_2 \cup B_1'\setminus\{b\}, B_2')$ is a split of a skew-partition $(A,B)$ in $T$.  The pair $(A_2', B_2')$ is
balanced in $T$ because it is balanced in $T_X$.  Hence,
by~\ref{onepair}, $T$ admits a balanced skew-partition, a
contradiction.   

The case when $T$ has a complement $2$-join follows by complementation
\bbox

\section{Handling basic trigraphs}
\label{sec:bas}

Our next goal is to compute maximum strong stable sets.  We need to
work in weighted trigraphs for the sake of induction. So, throughout
the remainder of the paper, by ``trigraph'' we mean a trigraph with
weights on the vertices.  Weights are numbers from $K$ where $K$ is
either the set $\mathbb{R}_+$ of non-negative real numbers or the set
$\mathbb{N}_+$ of non negative integers.  The statements of the
theorems will be true for $K= \mathbb{R}_+$ but the algorithms are to
be implemented with $K= \mathbb{N}_+$. Note that we view a trigraph
where no weight is assigned to the vertices as a weighted trigraph all
of whose vertices have weight~1.  Observe that a set of vertices of a
trigraph is a strong stable set if and only if it is a stable set of
its full realization.

\begin{theorem}
  \label{th:decBas}
  There is an $O(n^4)$ algorithm whose input is a trigraph and whose
  output is either the true statement ``$T$ is not basic'', or the
  name of a basic class in which $T$ is and the maximum weight of a
  strong stable set of $T$.
\end{theorem}

\Proof For each basic class, we provide an at most $O(n^4)$ time
algorithm that decides whether a trigraph $T$ belongs to the class,
and if so, computes a maximum weighted strong stable set.

For bipartite trigraphs, we construct the full realization $G$ of $T$.
It is easy to see that $T$ is bipartite if and only if $G$ is
bipartite, and deciding whether a graph is bipartite can be done in
linear time by the classical Breadth First Search.  If $T$ is
bipartite, a maximum weighted stable set of $G$ (which is a maximum
weighted strong stable set of $T$) can be computed in time $O(n^3)$,
see~\cite{schrijver:opticomb}.

For complements of bipartite trigraph, we proceed similarly: we first
take the complement $\overline{T}$ of the input trigraph $T$, and then
recognize whether the full realization of $\overline{T}$ is
bipartite. We then compute the maximum weighted clique in
$G^{\overline{T}}_{\emptyset}$. All this can clearly be done in ${\cal
  O}(n^2)$ time.

For line trigraphs, we compute the full realization $G$, and test
whether $G$ is a line graph of a bipartite graph by a classical
algorithm from~\cite{lehot:root} or~\cite{roussopoulos:linegraphe}.
Note that these algorithms also provide a graph $R$ such that
$G=L(R)$.  In time $O(n^3)$ we can check that every clique of size at
least~3 in $T$ is a strong clique so we can decided whether $T$ is a
line trigraph.  If so, a maximum stable set in $G$ can be computed in
time $O(n^3)$ by computing a maximum weighted matching
(see~\cite{schrijver:opticomb}) in a bipartite graph $R$ such that $G=
L(R)$.

For complements of line trigraphs, we proceed similarly for the
recognition except that we work with the full realization of $\overline{T}$.  And
computing a maximum weighted strong stable set is easy: compute the
full realization $G$ of $T$, then compute a bipartite graph $R$ such
that $G = \overline{L(R)}$ (this exists because by~\ref{linetrig},
line trigraphs are closed under taking realizations) and compute a
maximum weighted stable set in $G$ (note that such a set is an
inclusion-wise maximal set of pairwise adjacent edges in $R$, and there are
linearly many such sets).  This is a maximum weighted strong stable
set in~$T$.

For doubled trigraphs, the situation is slightly more complicated,
because we do not know how to rely on classical results.  But for one
who starts from scratch (with no knowledge of matching theory for
instance), they are in fact the easiest basic graphs to handle.  To
decide whether a graph $G$ is doubled, we may use the list of
minimally non-doubled graphs described in~\cite{alexeevFK:doubled}.
This list is made of 44 graphs on at most 9 vertices, so it yields an
$O(n^9)$ time recognition algorithm.  We propose here something
faster, and which also works for trigraphs.

If a partition $(X,Y)$ of the vertices of a trigraph is given,
deciding whether it is good can be done by a brute force checking of
all items from the definition in time $O(n^2)$.  And if an edge $ab$
from $T|X$ is given, reconstructing the good partition is easy: all
the vertices strongly antiadjacent to $a$ and $b$ go into $X$, and all
the vertices strongly adjacent to at least one of $a$ or $b$ go into
$Y$.  So, by checking all edges $uv$, one can guess one that is in
$T|X$, then reconstruct $(X, Y)$, and therefore test in time $O(n^4)$
whether a trigraph $T$ has a good partition $(X, Y)$ such that $X$
contains at least one edge.  Similarly, one can test in time $O(n^4)$
whether a trigraph $T$ has a good partition $(X, Y)$ such that $Y$
contains at least one antiedge.  We are left with the recognition of
doubled trigraphs such that all good partitions are made of one strong
stable set and one strong clique.  These are in fact graphs (there is
no switchable pair), and are known as \emph{split graphs} (in fact,
double split graphs were named after split graphs).  They can be
recognized in linear time, see~\cite{hammerS:split} where it it shown
that by looking at the degrees, one can easily output a partition of a
graph into a clique and a stable set, if any such partition exists.

Now, we know that $T$ is a doubled graph, and we look for a maximum
weighted strong stable set in $T$.  To do so, we compute the full
realization $G$ of~$T$.  So, by~\ref{bicotri} , $G$ is a doubled graph,
and in fact, $(X, Y)$ is good partition for~$G$.  We then compute a
maximum weighted stable set in $G|X$ (that is bipartite), in $G|Y$
(that is complement of bipartite), and all stable sets made of a
vertex from $Y$ together with its non-neighbors in $X$.  One of these
is a maximum weighted stable set of $G$, and so a strong one in $T$.
\bbox

\section{Keeping track of $\alpha$}
\label{sec:decAlpha}

In this section, we define several blocks of decompositions that allow
us to compute maximum strong stable sets.  From here on, $\alpha (T)$
denotes the weight of a maximum weighted strong stable set of $T$.

In what follows, $T$ is a trigraph from~$\cal F$ with no balanced
skew-partition, $X$ is an fragment of $T$ and $Y = V(T) \setminus X$ (so
$Y$ is also fragment of $T$).  To compute $\alpha (T)$, it is not enough to
consider the blocks $T_X$ and $T_Y$ (as defined in
Section~\ref{sec:BlockDec}) separately.  Instead, we need to enlarge
$T_Y$ slightly, to encode information from $X$. In this section, we define four different kinds
of gadgets, named $T_{Y, 1}$, \dots, $T_{Y, 4}$ and for $i=1, \dots,
4$, we prove that $\alpha(T)$ may easily be computed from
$\alpha(T_{Y, i})$.  We sometimes have to define different gadgets for
handling the same situation.  This is because in
Section~\ref{sec:computeAlpha} (namely to prove~\ref{expBas}), we need
that gadgets preserve being basic, and depending on the basic class
under consideration, we need to use different gadgets.  Note that the
gadgets are not class-preserving (some of them introduce balanced
skew-partitions).  In this section, this is not a problem, but in the
next section, this makes things a bit more complicated.

\subsection{Complement $2$-join}
\label{ss:c2j}

If $(X, Y)$ is a proper complement $2$-join of $T$ then let $X_1=X$,
$X_2=Y$, and let $(A_1,B_1,C_1,A_2,B_2,C_2)$ be a split of $(X_1,
X_2)$.  We build the gadget $T_{Y,1}$ as follows.  We
start with $T |Y$. We then add two new \emph{marker vertices} $a$,
$b$, such that $a$ is strongly complete to $B_2\cup C_2$, $b$ is
strongly complete to $A_2 \cup C_2$ and $ab$ is a strong edge.  We give
weights $\alpha_A = \alpha(T|A_1)$ and $\alpha_B = \alpha(T|B_1)$ to
$a$ and $b$ respectively.  We set $\alpha_X = \alpha(T|X)$.

\begin{lemma}
  \label{alphaC2J}
  If $(X, Y)$ is a proper complement $2$-join of $T$, then $T_{Y, 1}$
  is Berge and $\alpha(T) = \max (\alpha(T_{Y, 1}), \alpha_X)$.
\end{lemma}
\Proof Since $T_{Y, 1}$ is a semirealization of an induced subtrigraph of
the block $T_Y$ as defined in Section~\ref{sec:BlockDec}, it is clearly
Berge by~\ref{l:stayBerge}.

Let $Z$ be a maximum weighted strong stable set in $T$. If $Z
\cap X_1 = \emptyset$, then $Z$ is also a strong stable set in $T_{Y,
  1}$, so $\alpha(T) \leq \alpha(T_{Y, 1}) \leq \max (\alpha(T_{Y,
  1}), \alpha_X)$.  If $Z \cap A_1 \neq \emptyset$ and $Z \cap (B_1
\cup C_1) = \emptyset$, then $\{a_1\} \cup (Z\cap X_2)$ is a strong
stable set in $T_{Y, 1}$ of weight $\alpha(T)$, so $\alpha(T) \leq
\alpha(T_{Y, 1}) \leq \max (\alpha(T_{Y, 1}), \alpha_X)$.  If $Z \cap
B_1 \neq \emptyset$ and $Z \cap (A_1 \cup C_1) = \emptyset$, then
$\{b_1\} \cup (Z\cap X_2)$ is a strong stable set in $T_{Y, 1}$ of
weight $\alpha(T)$, so $\alpha(T) \leq \alpha(T_{Y, 1}) \leq \max
(\alpha(T_{Y, 1}), \alpha_X)$.  If $Z \cap (A_1 \cup C_1) \neq
\emptyset$ and $Z \cap (B_1 \cup C_1) \neq \emptyset$, then $\alpha(T)
= \alpha_X$, so $\alpha(T) \leq \max (\alpha(T_{Y, 1}), \alpha_X)$.
In all cases, we proved that $\alpha(T) \leq \max (\alpha(T_{Y, 1}),
\alpha_X)$.

Conversely, let $\alpha = \max (\alpha(T_{Y, 1}), \alpha_X)$.  If
$\alpha = \alpha_X$, then by considering any maximum strong stable set
of $T| X_1$, we see that $\alpha = \alpha_X \leq \alpha(T)$.  So we
may assume that $\alpha = \alpha(T_{Y, 1})$ and let $Z$ be a maximum
weighted strong stable set in $T_{Y, 1}$.  If $a \notin Z$ and $b
\notin Z$, then $Z$ is also a strong stable set in $T$, so $\alpha
\leq \alpha(T)$.  If $a \in Z$ and $b \notin Z$, then $Z' \cup Z
\setminus \{a\}$, where $Z'$ is a maximum weighted strong stable in
$T|A_1$, is also a strong stable set in $T$ of same weight as $Z$, so
$\alpha \leq \alpha(T)$.  If $a \notin Z$ and $b \in Z$, then $Z' \cup
Z \setminus \{b\}$ where $Z'$ is a maximum weighted stable in $T|B_1$
is also a strong stable set in $T$ of same weight as $Z$, so $\alpha
\leq \alpha(T)$.  In all cases, we proved that $\alpha \leq
\alpha(T)$.  \bbox

\subsection{$2$-join}
\label{ss:2j}

In~\cite{nicolas.kristina:2-join}, an NP-hardness result is proved,
that suggests that the $2$-join is maybe not the most convenient tool to
compute maximum stable sets.  It seems that to use them, we really
need to take advantage of Bergeness in some way.  This is done here by
proving several inequalities.

If $(X, Y)$ is a $2$-join of $T$ then let $X_1=X$, $X_2=Y$ and let
$(A_1,$ $B_1,$ $C_1,$ $A_2,$ $B_2,$ $C_2)$ be a split of $(X_1, X_2)$.
We define $\alpha_{AC} = \alpha(T|({A_1 \cup C_1}))$, $\alpha_{BC} =
\alpha(T|(B_1 \cup C_1))$, $\alpha_{C}= \alpha(T|C_1)$ and $\alpha_{X}
= \alpha(T|X_1)$.  Let $w$ be the weight function on $V(T)$.  When $H$
is an induced subtrigraph of $T$, or a subset of $V(T)$, $w(H)$
denotes the sum of the weights of vertices in $H$.

\begin{lemma}
  \label{l:4cases}
  Let $S$ be a maximum weighted strong stable set of $T$. Then exactly
  one of the following holds:

  \begin{enumerate}
  \item\label{i:4c1} $S \cap A_1 \neq \emptyset$, $S \cap B_1 =
    \emptyset$, $S\cap X_1$ is a maximum weighted strong stable set of $T| (A_1
    \cup C_1)$ and $w(S \cap X_1) = \alpha_{AC}$;
  \item\label{i:4c2} $S \cap A_1 = \emptyset$, $S \cap B_1 \neq
    \emptyset$, $S\cap X_1$ is a maximum weighted strong stable set of $T| (B_1
    \cup C_1)$ and $w(S \cap X_1) = \alpha_{BC}$;
  \item\label{i:4c3} $S \cap A_1 = \emptyset$, $S \cap B_1 =
    \emptyset$, $S\cap X_1$ is a maximum weighted strong stable set of
    $T|C_1$ and $w(S \cap X_1) = \alpha_{C}$;
  \item\label{i:4c4} $S \cap A_1 \neq \emptyset$, $S \cap B_1 \neq
    \emptyset$, $S\cap X_1$ is a maximum weighted strong stable set of
    $T|X_1$ and $w(S \cap X_1) = \alpha_{X}$.
  \end{enumerate}
\end{lemma}

\Proof
  Follows directly from the definition of a $2$-join.
\bbox

We need several inequalities that say more about how strong stable
sets and $2$-joins overlap.  These lemmas are proved
in~\cite{nicolas.kristina:2-join} in the context of graphs.  The
proofs are the same for trigraphs, but for the sake of
completeness, we rewrite them.

\begin{lemma}
  \label{l:ineqbasic}
  $0 \leq \alpha_{C} \leq \alpha_{AC}, \alpha_{BC} \leq \alpha_{X}
  \leq \alpha_{AC}+\alpha_{BC}$.
\end{lemma}

\Proof The inequalities $0 \leq \alpha_{C} \leq \alpha_{AC},
\alpha_{BC} \leq \alpha_{X}$ are trivially true. Let $D$ be a maximum
weighted strong stable set of $T| X_1$.  We have:
  $$
  \alpha_{X} = w(D) = w(D\cap A_1) + w(D\cap (C_1 \cup B_1)) \leq
  \alpha_{AC} + \alpha_{BC}.
  $$
\bbox

\begin{lemma}
  \label{l:ineqOdd}
  If $(X_1, X_2)$ is an odd $2$-join of $T$, then $\alpha_{C}+\alpha_{X}
  \leq \alpha_{AC}+\alpha_{BC}$.
\end{lemma}

\Proof Let $D$ be a strong stable set of $T | X_1$ of weight
$\alpha_{X}$ and $C$ a strong stable set of $T| C_1$ of weight $\alpha_{C}$.
In the bipartite trigraph $T | (C\cup D)$, we denote by $Y_A$
(resp.\ $Y_B$) the set of those vertices of $C\cup D$ for which there
exists a path in $T | (C \cup D)$ joining them to some vertex of
$D\cap A_1$ (resp.\ $D \cap B_1$).  Note that from the definition, $D
\cap A_1 \subseteq Y_A$, $D \cap B_1 \subseteq Y_B$ and there are no edges 
between $Y_A \cup Y_B$ and $(C\cup D)\setminus (Y_A \cup Y_B)$.  We claim that
$Y_A \cap Y_B= \emptyset$, and $Y_A$ is strongly anticomplete to $Y_B$. 
Suppose not. then there exists a  path $P$  in $T | (C\cup D)$ from a vertex of 
$D \cap A_1$ to a vertex of $D \cap B_1$.  We may assume that $P$ is minimal 
with respect to this property, and so the  interior of $P$ is in $C_1$; 
consequently $P$ is of even length because $T | (C\cup D)$ is bipartite.  
This contradicts the assumption that $(X_1, X_2)$ is odd.  Now we set:

  \begin{itemize}
  \item $Z_A = (D \cap Y_A) \cup (C \cap Y_B) \cup (C \setminus (Y_A \cup
    Y_B))$;
  \item $Z_B = (D \cap Y_B) \cup (C \cap Y_A) \cup (D \setminus (Y_A \cup Y_B)$.
  \end{itemize}

  From all the definitions and properties above, $Z_A$ and $Z_B$ are
  strong stable sets and $Z_A \subseteq A_1 \cup C_1$ and $Z_B
  \subseteq B_1 \cup C_1$.  So, $\alpha_{C}+\alpha_{X} = w(Z_A) +
  w(Z_B) \leq \alpha_{AC}+\alpha_{BC}$.  \bbox

\begin{lemma}
  \label{l:ineqEven}
  If $(X_1, X_2)$ is an even $2$-join of $T$, then
  $\alpha_{AC}+\alpha_{BC} \leq \alpha_{C}+\alpha_{X}$.
\end{lemma}

\Proof Let $A$ be a strong stable set of $T| (A_1 \cup C_1)$ of weight
$\alpha_{AC}$ and $B$ a strong stable set of $T | (B_1 \cup C_1)$ of
weight $\alpha_{BC}$.  In the bipartite trigraph $T | (A\cup B)$, we
denote by $Y_A$ (resp.\ $Y_B$) the set of those vertices of $A\cup B$
for which there exists a path $P$ in $T | (A \cup B)$ joining them to
a vertex of $A \cap A_1$ (resp.\ $B \cap B_1$).  Note that from the
definition, $A \cap A_1 \subseteq Y_A$, $B \cap B_1 \subseteq Y_B$,
and $Y_A\cup Y_B$ is strongly anticomplete to $(A\cup B)\setminus (Y_A
\cup Y_B)$.  We claim that $Y_A \cap Y_B = \emptyset$ and $Y$ is
strongly anticomplete to $Y_B$. Suppose not, then there is a path $P$
in $T | (A\cup B)$ from a vertex of $A \cap A_1$ to a vertex of $B
\cap B_1$.  We may assume that $P$ is minimal with respect to this
property, and so the interior of $P$ is in $C_1$; consequently it is
of odd length because $T (A\cup B)$ is bipartite.  This contradicts
the assumption that $(X_1, X_2)$ is even.  Now we set:

  \begin{itemize}
  \item $Z_D = (A \cap Y_A) \cup (B \cap Y_B) \cup (A \setminus (Y_A \cup Y_B))$;
  \item $Z_C = (A \cap Y_B) \cup (B \cap Y_A) \cup (B \setminus (Y_A \cup Y_B))$.
  \end{itemize}

  From all the definitions and properties above, $Z_D$ and $Z_C$ are
  strong stable sets and $Z_D \subseteq X_1$ and $Z_C \subseteq C_1$.
  So, $\alpha_{AC}+\alpha_{BC} = w(Z_C) + w(Z_D) \leq
  \alpha_{C}+\alpha_{X}$.  \bbox

We are now ready to build the gadgets. 

If $(X_1,X_2)$ is a proper odd $2$-join of $T$, then we build the
gadget $T_{Y,2}$ as follows. We start with $T |Y$. We then add four
new \emph{marker vertices} $a$, $a'$, $b$, $b'$, such that $a$ and
$a'$ are strongly complete to $A_2$, $b$ and $b'$ are strongly
complete to $B_2$, and $ab$ is a strong edge.  We give weights
$\alpha_{AC} + \alpha_{BC} - \alpha_{C} - \alpha_{X}$, $\alpha_X -
\alpha_{BC}$, $\alpha_{AC} + \alpha_{BC} - \alpha_{C} - \alpha_{X}$
and $\alpha_X - \alpha_{AC}$ to $a$, $a'$, $b$ and $b'$ respectively.
Note that by~\ref{l:ineqbasic} and~\ref{l:ineqOdd}, all the weights
are non-negative.

We define another gadget of decomposition $T_{Y,3}$ for the same
situation, as follows. We start with $T |Y$. We then add three new
\emph{marker vertices} $a$, $a'$, $b$, such that $a$ and $a'$ are
strongly complete to $A_2$, $b$ is strongly complete to $B_2$, and
$a'a$ and $ab$ are strong edges.  We give weights $\alpha_{AC} -
\alpha_{C}$, $\alpha_{X} - \alpha_{BC}$ and $\alpha_{BC} - \alpha_{C}$
to $a$, $a'$ and $b$ respectively.  Note that by~\ref{l:ineqbasic},
all the weights are non-negative.

\begin{lemma}
  \label{l:oddblock3}
  If $(X, Y)$ is a proper odd $2$-join of $T$, then $T_{Y, 2}$ and
  $T_{Y, 3}$ are Berge, and $\alpha(T) = \alpha(T_{Y, 2}) + \alpha_C =
  \alpha(T_{Y, 3}) + \alpha_C$.
\end{lemma}

\Proof Suppose that $T_{Y, 2}$ contains an odd hole $H$.  Since an odd
hole has no strongly dominated vertex, it contains at most one of $a,
a'$ and at most one of $b, b'$.  Hence, $H$ is an odd hole of some
semirealization of the block $T_Y$ (as defined in
Section~\ref{sec:BlockDec}). This contradicts~\ref{l:stayBerge}.
Similarly, $T_{Y, 2}$ contains no odd antihole, and therefore, it is
Berge.  The proof that $T_{Y, 3}$ is Berge is similar.

Let $Z$ be a strong stable set in $T$ of weight $\alpha(T)$.  We build
a strong stable set in $T_{Y, 2}$ by adding to $Z \cap X_2$ one the
following (according to the outcome of~\ref{l:4cases}): $\{a, a'\}$,
$\{b, b'\}$, $\emptyset$, or $\{a, a', b'\}$.  In each case, we obtain
a strong stable set of $T_{Y, 2}$ with weight $\alpha(T) - \alpha_C$.
This proves that $\alpha(T) \leq \alpha(T_{Y, 2}) + \alpha_C$.

Conversely, let $Z$ be a stable set in $T_{Y, 2}$ with weight
$\alpha(T_{Y, 2})$.  We may assume that $Z \cap \{a, a', b, b'\}$ is
one of $\{a, a'\}$, $\{b, b'\}$, $\emptyset$, or $\{a, a', b'\}$, and
respectively to these cases, we construct a strong stable set of $T$
by adding to $Z\cap X_2$ a maximum weighted strong stable set of the
following: $T|(A_1 \cup C_1)$,  $T|(B_1 \cup C_1)$, $T|C_1$, or
$T|X_1$.  We obtain a strong stable set in $T$ with weight
$\alpha(T_{Y, 2}) + \alpha_C$, showing that  $\alpha(T_{Y, 2}) +
\alpha_C \leq \alpha(T)$. This completes the proof for $T_{Y, 2}$. 

Let us now prove the equality for $T_{Y, 3}$.  Let $Z$ be a strong
stable set in $T$ of weight $\alpha(T)$.  We build a strong stable set
in $T_{Y, 3}$ by adding to $Z \cap X_2$ one the following (according
to the outcome from~\ref{l:4cases}): $\{a\}$, $\{b\}$, $\emptyset$, or
$\{a', b\}$.  In each case, we obtain a strong stable set of $T_{Y,
  3}$ with weight $\alpha(T) - \alpha_C$.  This proves that $\alpha(T)
\leq \alpha(T_{Y, 3}) + \alpha_C$.

Conversely, let $Z$ be a stable set in $T_{Y, 3}$ with weight
$\alpha(T_{Y, 3})$.  By~\ref{l:ineqOdd}, $\alpha_{AC} - \alpha_C \geq
\alpha_X - \alpha_{BC}$, so we may assume that $Z \cap \{a, a', b\}$
is one of $\{a\}$, $\{b\}$, $\emptyset$, or $\{a', b\}$, and
respectively to these cases, we construct a strong stable set of $T$
by adding to $Z\cap X_2$ a maximum weighted strong stable set of the
following: $T|(A_1 \cup C_1)$, $T|(B_1 \cup C_1)$, $T|C_1$, or
$T|X_1$.  We obtain a strong stable set in $T$ with weight
$\alpha(T_{Y, 3}) + \alpha_C$, showing that $\alpha(T_{Y, 3}) +
\alpha_C \leq \alpha(T)$. This completes the proof for $T_{Y, 3}$.
\bbox

If $(X_1,X_2)$ is a proper even $2$-join of $T$ and $X=X_1$, $Y=X_2$,
then let $(A_1,B_1,C_1,A_2,B_2,C_2)$ be a split of $(X_1, X_2)$. We
build the gadget $T_{Y, 4}$ as follows.  We start with
$T |Y$.  We then add three new \emph{marker vertices} $a$, $b$, $c$
such that $a$ is strongly complete to $A_2$, $b$ is strongly complete
to $B_2$, and $c$ is strongly adjacent to $a, b$ and has no other
neighbors.  We give weights $\alpha_X - \alpha_{BC}$, $\alpha_X -
\alpha_{AC}$, and $\alpha_{X} + \alpha_{C} - \alpha_{AC} -
\alpha_{BC}$ to $a$, $b$ and $c$ respectively.  Note that
by~\ref{l:ineqbasic} and~\ref{l:ineqEven}, these weights are non-negative.

\begin{lemma}
  \label{l:evenblock4}
  If $(X, Y)$ is a proper even $2$-join of $T$, then $T_{Y, 4}$ is
  Berge and $\alpha(T) = \alpha(T_{Y, 4}) + \alpha_{AC} + \alpha_{BC}
  - \alpha_{X}$.
\end{lemma}
\Proof 
Clearly, $T_{Y, 4}$ is Berge, because it is a semirealization of the
block $T_Y$ as defined in Section~\ref{sec:BlockDec}, which is Berge
by~\ref{l:stayBerge}.  

Let $Z$ be a strong stable set in $T$ of weight $\alpha(T)$.
We build a strong stable set in $T_{Y, 4}$ by adding to $Z \cap X_2$
one the following (according to the outcome of~\ref{l:4cases}):
$\{a\}$, $\{b\}$, $\{c\}$, or $\{a, b\}$.  In each case, we obtain a
strong stable set of $T_{Y, 4}$ with weight $\alpha(T) - (\alpha_{AC}
+ \alpha_{BC} - \alpha_{X})$.  This proves that $\alpha(T) \leq
\alpha(T_{Y, 4}) + \alpha_{AC} + \alpha_{BC} - \alpha_{X}$.

Conversely, let $Z$ be a strong stable set in $T_{Y, 4}$ with weight
$\alpha(T_{Y, 4})$.  We may assume that $Z \cap \{a, b, c\}$ is one of
$\{a\}$, $\{b\}$, $\{c\}$, or $\{a, b\}$, and respectively to these
cases, we construct a strong stable set of $T$ by adding to $Z\cap
X_2$ a maximum weighted strong stable set of the following: $T|(A_1
\cup C_1)$, $T|(B_1 \cup C_1)$, $T|C_1$, or $T|X_1$.  We obtain a
strong stable set in $T$ with weight $\alpha(T_{Y, 4}) + \alpha_{AC} +
\alpha_{BC} - \alpha_{X}$, showing that $\alpha(T_{Y, 4}) +
\alpha_{AC} + \alpha_{BC} - \alpha_{X} \leq \alpha(T)$.\bbox

\section{Computing $\alpha$}
\label{sec:computeAlpha}

We are ready to describe our main algorithm, that computes a maximum
weighted stable set.  The main difficulty is that blocks of
decompositions as defined in Section~\ref{sec:BlockDec} have to be used
in order to stay in the class, while gadgets as defined in
Section~\ref{sec:decAlpha} have to be used for computing $\alpha$.
Our idea is to use blocks in a first stage, and to replace them by
gadgets in a second stage.  To transform a block into a gadget (this
operation is called an \emph{expansion}), one needs to erase a
switchable component, and to replace it by some vertices with the
appropriate weights.  Two kinds of information are needed.  The first
one is the type of decomposition that is originally used and the
weights; this information is encoded into what we call a
\emph{prelabel}.  The second one is the type of basic class in which
the switchable component ends up (because not all gadgets preserve
being a basic class); this information is encoded into what we call a
\emph{label}.  Note that the prelabel is known right after decomposing
a trigraph, while the label becomes known much later, when the
decomposition is fully processed.  Let us make all this formal.

Let $S$ be a switchable component of a trigraph $T$ from $\cal F$.  A
\emph{prelabel} for $S$ is one of the following: 

\begin{itemize}
\item $($``Complement odd $2$-join", $\alpha_A, \alpha_B, \alpha_X)$
  where $\alpha_A$, $\alpha_B$ and $\alpha_X$ are integers, if $S$ is
  a switchable pair.

\item  $($``Odd $2$-join", $\alpha_{AC},
  \alpha_{BC}, \alpha_C, \alpha_X)$ where $\alpha_{AC}$,
  $\alpha_{BC}$, $\alpha_C$ and $\alpha_X$ are integers,
if $S$ is switchable pair and no vertex of $T$ is complete to $S$.

\item  $($``Complement even $2$-join", $\alpha_A, \alpha_B, \alpha_X)$ where
  $\alpha_A$, $\alpha_B$ and $\alpha_X$ are integers, if $S$ is a heavy 
  component.

\item $($``Even $2$-join",
  $\alpha_{AC}, \alpha_{BC}, \alpha_C, \alpha_X)$ where $\alpha_{AC}$,
  $\alpha_{BC}$, $\alpha_C$ and $\alpha_X$ are integers, if $S$ is a light 
  component.

\end{itemize} 

We remark that certain types of switchable components are ``eligible''
for both the first and second type of prelabel.

A prelabel should be thought of as ``the decomposition from which the
switchable component has been built''.  When $T$ is a trigraph and
$\cal S$ is a set of switchable components of $T$, a
\emph{prelabeling} for $(T, {\cal S})$ is a function that associates
to each $S\in {\cal S}$ a prelabel.  It is important to notice that
$\cal S$ is just a set of switchable component, so that some
switchable components may have no prelabel. 

What follows is slightly ambiguous when we talk about ``the basic
class containing the trigraph '', because some trigraphs may be
members of several basic classes (typically, small trigraphs, complete
trigraphs, independent trigraphs and a few others).  But this is not a
problem; if a trigraph belongs to several basic classes, our algorithm
chooses one such class arbitrarily, and the output is correct.  We
choose not to make this too formal and heavy, so this is not mentioned
explicitly in the descriptions of the algorithms.  For doubled graphs,
there is one more ambiguity.  Let $T$ be a doubled graph, and $(X, Y)$
 a good partition of $T$.  A switchable pair $uv$ of $T$ is a
\emph{matching pair} if $u, v \in X$ and an \emph{antimatching pair}
if $u, v \in Y$.  In some small degenerate cases, a switchable pair of
a doubled graph may be a matching and an antimatching pair according
to the good partition under consideration, but once a good partition
is fixed, there is no ambiguity.  Again, this is not a problem: when a
pair is ambiguous, the algorithm chooses arbitrarily one particular
good partition.

Let $S$ be a switchable component of a trigraph from $\cal F$.  A
\emph{label} for $S$ is a pair $L' = (L, N)$ such that $L$ is a
prelabel and $N$ is one of the following: ``bipartite'', ``complement
of bipartite'', ``line'', ``complement of line'',
``doubled-matching'', ``doubled-antimatching''.  We say that $L'$
\emph{extends}~$L$.  The tag added to extend a prelabel of a
switchable component $S$ should be thought of as ``the basic class in
which $S$ ends up when the trigraph is fully decomposed''.  When $T$
is a trigraph and $\cal S$ is a set of switchable components of $T$, a
\emph{labeling} for $(T, {\cal S})$ is a function that associates to
each $S\in {\cal S}$ a label.  Under these circumstances we say that
$T$ is {\em labeled}.  As with prelabels, switchable components not in
$\cal S$ receive no label.

Let $T$ be a labeled trigraph, $\cal S$ a set of switchable components
of $T$ and $\cal L$ a labeling for $(T, {\cal S})$.  The
\emph{expansion} of $(T, {\cal S}, {\cal L})$ is the trigraph obtained
from $T$ after performing for each $S \in {\cal S}$ with label $L$ the
following operation:

\begin{enumerate}
\item If $L = (($``Complement odd $2$-join'', $\alpha_A, \alpha_B,
  \alpha_X), N)$ for some $N$ (so $S$ is a switchable pair $ab$):
  transform $ab$ into a strong edge, give weight $\alpha_A$ to 
  $a$ and weight $\alpha_B$ to $b$.

\item If  $L = (($``Odd $2$-join'', $\alpha_{AC}, \alpha_{BC},
  \alpha_C, \alpha_X), N)$ for some $N$ (so $S$ is a switchable pair
  $ab$): transform $ab$ into a strong edge, and:
  \begin{itemize}  
  \item If $N$ is equal to one of ``bipartite'', ``complement of
    line'', or ``doubled-matching'', then add a vertex $a'$, a vertex
    $b'$, make $a'$ strongly complete to $N(a)\setminus \{b\}$, make
    $b'$ strongly complete to $N(b)\setminus \{a\}$, and give weights
    $\alpha_{AC} + \alpha_{BC} - \alpha_{C} - \alpha_{X}$, $\alpha_X -
    \alpha_{BC}$, $\alpha_{AC} + \alpha_{BC} - \alpha_{C} -
    \alpha_{X}$ and $\alpha_X - \alpha_{AC}$ to $a$, $a'$, $b$ and
    $b'$ respectively.
  \item If $N$ is equal to one of ``complement of bipartite'',
    ``line'' or ``doubled-antimatching'', then add a vertex $a'$, make
    $a'$ strongly complete to $\{a\} \cup N(a) \setminus \{b\}$, and
    give weights $\alpha_{AC} - \alpha_{C}$, $\alpha_{X} -
    \alpha_{BC}$ and $\alpha_{BC} - \alpha_{C}$ to $a$, $a'$ and $b$
    respectively.
  \end{itemize}

\item If $L = (($``Complement even $2$-join'', $\alpha_A, \alpha_B,
  \alpha_X), N)$ for some $N$ (so $S$ is made of two switchable pairs
  $ac$ and $cb$ and $c$ is heavy): delete the vertex $c$, and give weight 
  $\alpha_A$ to $a$ and   weight $\alpha_B$ to $b$.

\item If  $L = (($``Even $2$-join'', $\alpha_{AC}, \alpha_{BC},
  \alpha_C, \alpha_X), N)$ for some $N$ (so $S$ is made of two
  switchable pairs $ac$ and $cb$, and $c$ is light): transform  $ac$
  and $cb$ into strong edges, and give weights $\alpha_X -
  \alpha_{BC}$, $\alpha_X - \alpha_{AC}$, and $\alpha_{X} + \alpha_{C}
  - \alpha_{AC} - \alpha_{BC}$ to $a$, $b$ and $c$ respectively.

\end{enumerate}

The expansion should be thought of as ``what is obtained if one uses a
gadget as defined in Section~\ref{sec:decAlpha} instead of a
block of decomposition as defined in Section~\ref{sec:BlockDec}''.  

\begin{theorem}
  \label{expBas}
  Suppose that $T$ is trigraph that is in a basic class  with name $N$,
  $\cal S$ is a set of switchable components of $T$ and $\cal L$ is a
  labeling for $T$ such that for all $S\in {\cal S}$ with label $L$,
  one of the following holds:
  \begin{itemize}
  \item $L = (\dots, N)$ where $N$ is ``bipartite'', ``complement of
    bipartite'', ``line" or "complement of line''; or
  \item $N =$ ``doubled'', $S$ is a matching pair of $T$ and $L =
   (\dots,$ ``doubled-matching''$)$; or 
 \item $N =$ ``doubled'', $S$ is an antimatching pair of $T$ and $L =
   (\dots,$ ``doubled-antimatching''$)$. 
 \end{itemize}

 Then the expansion of $(T, {\cal S}, {\cal L})$ is a basic trigraph.
\end{theorem}

\Proof From our assumptions, $T$ is basic.  So, it is enough to prove
that expanding one switchable component $S$ preserves being basic, and
the result then follows by induction on $|\cal S|$. Let $T'$ be the
expansion.  For several cases from the definition of expansions
(namely items 1, 3 and 4), expanding just means possibly transforming
some switchable pairs into strong edges, and possibly deleting a
vertex.  From~\ref{l:presBas}, this preserves being basic. Hence, in
the argument below, we just study item 2 from the definition of
expansions, and thus we may assume that $S$ is a switchable pair
$ab$. 

It is easy to check that expansion as defined in item~2 preserves
being bipartite and being complement bipartite; so if $N \in
\{$``bipartite'', ``complement of bipartite''$\}$, then we are done.

Suppose that $N=$``line'', and so $T$ is a line trigraph.  Let $G$ be
the full realization of $T$, and $R$ a bipartite graph such that $G =
L(R)$.  So $a$ is an edge $x_a y_a$ in $R$, and $b$ is an edge $y_a
x_b$.  Since $T$ is a line trigraph, it follows that every clique of
size at least $3$ in $T$ is a strong clique, and so $a$ and $b$ have
no common neighbors in $T$. Therefore all the neighbors of $a$ except
$b$ are edges incident with $x_a$, and not with $y_a$.  Let $R'$ be
the graph obtained from $R$ by adding a pendant edge $e$ at $x_a$.  We
observe that $L(R')$ is isomorphic to the full realization of $T'$
(the edge $e$ yields the new vertex $a'$), and therefore $T'$ is a
line trigraph.

Next suppose that $N=$``complement of line'', so $T$ is the complement
of a line trigraph.  Since every clique of size at least $3$ in
$\overline{T}$ is a strong clique, it follows that $V(T) = N(a) \cup
N(b)$. Assume that there exist $u,v \in N(a) \setminus N(b)$ such that
$u$ is adjacent to $v$. Since $\overline{T}$ is a line trigraph, and
if $uv$ is a semiedge then $\{u,v,b\}$ is a clique of size $3$ in
$\overline{T}$, it follows that $u$ is strongly adjacent to $v$ in
$T$. Let $R$ be a bipartite graph such that the full realization of
$\overline{T}$ is $L(R)$.  Then in $R$ no two of the edges $u,v,a$
share and end, and yet $b$ shares an end with all three of them, a
contradiction.  This proves that $N(a) \setminus N(b)$ (and
symmetrically $N(b) \setminus N(a)$) is a strongly stable set in
$T$. As $N(a) \cap N(b)=\emptyset$, then $T$ is bipartite, and so a
previous argument shows that $T'$ is basic.

So we may assume that $T$ is a doubled trigraph with a good partition
$(X, Y)$. If $S=ab$ is a matching pair of $T$, then adding the
vertices $a', b'$ to $X$, produces a good partition of $T'$.  If
$S=ab$ is an antimatching pair of $T$, then adding the vertex $a'$ to
$Y$ produces a good partition of $T'$. Thus in all cases $T'$ is basic
and the theorem holds.  \bbox

Let $T$ be a trigraph, $\cal S$ a set of switchable components of $T$,
$\cal L$ a labeling of $(T, {\cal S})$ and $T'$ the expansion of $(T,
{\cal S}, {\cal L})$.  Let $X \subseteq V(T)$.  We define the  \emph{expansion
  $X'$ of $X$} as follows. Start with $X'=X$ and perform the following for  every $S \in \mathcal {S}$. 

\begin{enumerate}
\item If $L = (($``Complement odd $2$-join'', $\alpha_A, \alpha_B,
  \alpha_X), N)$ for some $N$ (so $S$ is a switchable pair $ab$), do not change $X'$.

\item If  $L = (($``Odd $2$-join'', $\alpha_{AC}, \alpha_{BC},
  \alpha_C, \alpha_X), N)$ for some $N$ (so $S$ is a switchable pair
  $ab$): 
  \begin{itemize}  
  \item If $N$ is equal to one of ``bipartite'', ``complement of
    line'', or ``doubled-matching'', do: 
    if $a\in X$ then add $a'$ to $X'$, and if $b \in X$ then add $b'$ to $X'$.

\item If $N$ is equal to one of ``complement of bipartite'',
    ``line'' or ``doubled-antimatching'', do:  
    if $a\in X$ then add $a'$ to $X'$.
\end{itemize}

\item If $L = (($``Complement even $2$-join'', $\alpha_A, \alpha_B,
  \alpha_X), N)$ for some $N$ (so $S$ is made of two switchable pairs
  $ac$ and $cb$ and $c$ is heavy), do: if $c \in X$, then remove $c$ from $X'$.

\item If $L = (($``Even $2$-join'', $\alpha_{AC}, \alpha_{BC},
  \alpha_C, \alpha_X), N)$ for some $N$ (so $S$ is made of two
  switchable pairs $ac$ and $cb$, and $c$ is light), do not change $X'$.

\end{enumerate}

\begin{theorem}
  \label{l:expDec}
  With the notation as above, if $(X_1, X_2)$ is a proper (complement)
  $2$-join of $T$ with split $(A_1, B_1, C_1, A_2, B_2, C_2)$, then
  $(X'_1, X'_2)$ is a proper (complement) $2$-join of $T'$ with split
  $(A'_1, B'_1, C'_1, A'_2, B'_2, C'_2)$, with same parity as $(X_1,
  X_2)$ (note that the notion of parity makes sense for $T'$, since
  $T'$ is Berge by~\ref{alphaC2J}, \ref{l:oddblock3} and~\ref{l:evenblock4}).
\end{theorem}

\Proof
Follows easily from the definitions. 
\bbox

\begin{theorem}
  \label{mainAlg}
  There exists an algorithm with the following specification.
  \begin{description}
  \item[Input:] A triple $(T, {\cal S}, {\cal L})$ such that $T$ is a
    trigraph in $\cal F$ with no balanced skew-partition, $\cal S$ is
    a set of switchable components of $T$ and $\cal L$ is a
    prelabeling for $(T, {\cal S})$.
  \item[Output:] A labeling ${\cal L}'$ for $(T, {\cal S})$ that
    extends $\cal L$, and a maximum weighted strong stable set of the
    expansion of $(T, {\cal S}, {\cal L}')$.
    \item[Running time:] $O(n^5)$ 
  \end{description}
\end{theorem}

\Proof We describe a recursive algorithm.  The first step of the
algorithm is to use~\ref{th:decBas} to check whether $T$ is basic.
Note that if $T$ is a doubled trigraph, the algorithm
from~\ref{th:decBas} also outputs which switchable pair is a
matching-pair, and which switchable pair is an antimatching pair.

Suppose first that $T$ is in a basic class with name~$N$ (this is the
case in particular when $|V(T)| = 1$).  We extend the prelabeling
$\cal L$ into a labeling $\cal L'$ as follows: if $N\neq$``doubled'',
then we append $N$ to every label and otherwise, for each $S\in {\cal
  S}$ with label $L$, we add ``doubled-matching'' (resp.\
``doubled-antimatching'') to $L$ when $S$ is a matching (resp.\
antimatching) pair.  It turns out that the labeling that we obtain
satisfies the requirements of~\ref{expBas}, so the expansion $T'$ of
$(T, {\cal S}, {\cal L'})$ is basic, and by running the algorithm
from~\ref{th:decBas} again for $T'$, we obtain a maximum weighted
strong stable set of $T'$ in time $O(n^4)$.  So, as claimed, we may
output a labeling ${\cal L}'$ for $(T, {\cal S})$ that extends $\cal
L$, and a maximum weighted strong stable set of the expansion of $(T,
{\cal S}, {\cal L}')$

Suppose now that $T$ is not basic.  Since $T$ is in $\cal F$ and has
no balanced skew-partition, by~\ref{structure}, we know that $T$ has a
$2$-join or the complement of a $2$-join.  In~\cite{ChHaTrVu:2-join},
an $O(n^4)$ time algorithm for computing a $2$-join in any input graph
is described.  In fact 2-joins as defined in~\cite{ChHaTrVu:2-join}
are sligthly different from the ones we use: they are not required to
satisfy the last item in our definition of a 2-join (this item ensures
to no side of the 2-join is a path of length exactly~2).  But the
method from Theorem~4.1 in~\cite{ChHaTrVu:2-join} shows how to handle
these kinds of requirements  with no additional time.  It is easy to
adapt this method to the detection of a $2$-join in a trigraph (also
Section~\ref{sec:end} of the present article gives a similar
algorithm).  So we can find the decomposition that we need in time
$O(n^4)$. We then compute the blocks $T_X$ and $T_Y$ as defined in
Section~\ref{sec:BlockDec}.  Note that every member of $\cal S$ is a
switchable pair of exactly one of $T_X$ or $T_Y$.  We call ${\cal
  S}_X$ (resp.\ ${\cal S}_Y$) the set formed by the members of $\cal
S$ that are in $T_X$ (resp.\ $T_Y$). Let $S$ be the marker switchable
component used to create the block $T_Y$.  Observe that for every $u
\in S$ there exists a vertex $v \in X$ such that $N_T(v)\cap
Y=N_{T_Y}(u)\cap Y$.  The same is true for $T_X$.  So, the prelabeling
$\cal L$ for $(T, {\cal S})$ naturally yields a prelabeling ${\cal
  L}_X$ for $(T_X, {\cal S}_X)$ and a prelabeling ${\cal L}_Y$ for
$(T_Y, {\cal S}_Y)$ (each $S\in {\cal S}_X$ receives the same prelabel
it has in $\cal L$, similarly for ${\cal S}_Y$).  In what follows,
\emph{the decomposition} refers to the decomposition that was used to
build $T_X$ and $T_Y$, (so one of ``complement odd $2$-join'',
``complement even $2$-join'', ``odd $2$-join'' or ``even $2$-join'')
and we use our standard notation for a split of the decomposition.

Up to symmetry, we may assume that $|V(T_X)|\leq
|V(T_Y)|$. By~\ref{l:stayBerge}, $T_X, T_Y$ are trigraphs from $\cal F$,
and by~\ref{stab}, they have no balanced skew-partition.

Let $S$ be the marker switchable component that was used to create
block $T_Y$.  We set ${\cal S}'_Y = {\cal S}_Y \cup \{S\}$.  We now
build a prelabeling ${\cal L}_Y$ for ${\cal S}'_Y$ as follows.  All
switchable components in ${\cal S}_Y$ keep the prelabel that they have
in $\cal S$.  The marker component $S$ receives the following
prelabel:
\begin{itemize}
\item If the decomposition is a complement odd $2$-join, then
  recursively compute
  $\alpha_A = \alpha(T_X|A_1)$, $\alpha_B = \alpha(T_X|B_1)$ and
  $\alpha_X = \alpha(T_X|X)$, and define the prelabel of $S$ as
  $($``Complement odd $2$-join'', $\alpha_A, \alpha_B, \alpha_X)$.
  Observe that in this case $|S|=2$.

\item If the decomposition is an odd $2$-join, then recursively compute $\alpha_{AC}
  = \alpha(T_X|({A_1 \cup C_1}))$, $\alpha_{BC} = \alpha(T_X|(B_1
  \cup C_1))$, $\alpha_{C}= \alpha(T_X|C_1)$ and $\alpha_{X} =
  \alpha(T_X|X)$ and define the prelabel of $S$ as $($``Odd $2$-join'',
  $\alpha_{AC}, \alpha_{BC}, \alpha_C, \alpha_X)$. Observe that in
  this case $|S|=2$ 
  and no vertex of $T'_Y \setminus S$ is strongly complete to $S$.

\item If the decomposition is a complement even $2$-join, then
  recursively compute
  $\alpha_A = \alpha(T_X|A_1)$, $\alpha_B = \alpha(T_X|B_1)$ and
  $\alpha_X = \alpha(T_X|X)$, and define the prelabel of $S$ as
  $($``Complement even $2$-join'', $\alpha_A, \alpha_B, \alpha_X)$.  
  Observe that in  this case $|S|=3$ and $S$ is light.

\item If the decomposition is an even $2$-join, then recursively compute
  $\alpha_{AC} = \alpha(T_X|({A_1 \cup C_1}))$, $\alpha_{BC} =
  \alpha(T_X|(B_1 \cup C_1))$, $\alpha_{C}= \alpha(T_X|C_1)$ and
  $\alpha_{X} = \alpha(T_X|X)$ and define the prelabel of $S$ as
  $($``Even $2$-join'', $\alpha_{AC}, \alpha_{BC}, \alpha_C, \alpha_X)$.
 Observe that in  this case $|S|=3$ and $S$ is heavy.

\end{itemize}

Now, $(T_Y, {\cal S}'_Y)$ has a prelabeling ${\cal L}_Y$.  We
recursively run our algorithm for $(T_Y, {\cal S}'_Y, {\cal L}_Y)$.

We obtain an extension ${\cal L}'_Y$ of ${\cal L}_Y$ and a maximum
weighted strong stable set of the expansion $T'_Y$ of $(T_Y, {\cal S}'_Y,
{\cal L}'_Y)$.

We use ${\cal L}'_Y$ to finish the construction of ${\cal L}'$, 
using for each $S \in {\cal S}_Y$ the same extension as we have in
${\cal L}'_Y$ for extending ${\cal L}_Y$.  Hence, now, we have an
extension ${\cal L}'$ of ${\cal L}$.  Let $T'$ be the expansion of
$(T, S, {\cal L}')$.

Observe now that by~\ref{l:expDec}, $T'_Y$ is precisely a gadget for
$T'$, as defined in Section~\ref{sec:decAlpha}.  Hence, $\alpha(T')$
may be recovered from $\alpha(T'_Y)$, as explained in one
of~\ref{alphaC2J}, \ref{l:oddblock3}, or \ref{l:evenblock4}.

Hence, the algorithm works  correctly when it returns ${\cal L}'$ and the 
maximum weight of a strong stable set that we have just computed.

\medskip

\noindent{\bf Complexity analysis:} By the way we construct our
blocks of decomposition, we have $|V(T_X)|-3+|V(T_Y)|-3\leq n$ and
by~\ref{2joinform}(\ref{size}) we have $6\leq |V(T_X)|,|V(T_Y)| \leq
n-1$. Recall that we have assumed that $|V(T_X)|\leq |V(T_Y)|$.

Let $T(n)$ be the complexity of our algorithm. For each kind of
decomposition we perform at most four recursive calls on the small
size, namely $T_X$, and one recursive call for the big side $T_Y$.
So we have $T(n) \leq d n^4$ when the graph is basic and otherwise
$T(n) \leq 4T(|V(T_X)|) + T(|V(T_Y)|) + d n^4$, where $d$ is the
constant arising from the complexity of finding a $2$-join or a
complement $2$-join and finding $\alpha$ in basic trigraphs.

We now prove that there exists a constant $c$ such that $T(n) \leq
cn^5$.  Our proof is by induction on $n$.  We show that there exists a
constant $N$ such that the induction step of our induction goes
through for all $n\geq N$ (this argument, and in particular $N$, does
not depend on $c$).  The base case our induction is therefore graphs
that are either basic or that have most $N$ vertices.  For them, $c$
clearly exists.

We write the proof of the induction step only when the decomposition
under consideration is an even $2$-join (possibly in the complement).
The proof for the odd $2$-join is similar.  We set $n_1 = |V(T_X)|$.  We
have $T(n) \leq 4T(n_1)+T(n+6-n_1)+dn^4$ for all $n_1$ and $n$
satisfying $\lfloor\frac{n}{2}\rfloor+3\geq n_1\geq 7$.

Let us define $f(n_1)=n^5 - 4n_1^5 - (n+6-n_1)^5 - dn^4$.  We show
that there exists a constant $N$ such that for all $n\geq N$ and all
$n_1$ such that $7\leq n_1 \leq \lfloor\frac{n}{2}\rfloor+3$,
$f(n_1)\geq 0$.  By the induction hypothesis, this proves our claim. A
simple computation yields:

\[
f'(n_1)=-20n_1^4+5(n+6-n_1)^4
\]

\[
f''(n_1)=-80n_1^3-20(n+6-n_1)^3
\]

Since $n+6-n_1$ is positive, we have $f''\leq 0$.  So, $f'$ is decreasing,
and it is easy to see that if $n$ is large enough, it is positive for
$n_1=7$ and negative for $n_1=\lfloor\frac{n}{2}\rfloor+3$.  Now $f$
is minimum for $n_1=7$ or $n_1=\lfloor\frac{n}{2}\rfloor+3$.  Since
$f(7)=n^5-(n-1)^5-P(n)$ where $P$ is a polynomial with $\deg(P)\leq
4$, if $n$ is large enough, then $f(7)$ is positive.  Also
$f(\lfloor\frac{n}{2}\rfloor+3)\leq n^5-5(\lceil\frac{n}{2}\rceil
+3)^5$. Again, if $n$ is large enough,
$f(\lfloor\frac{n}{2}\rfloor+3)$ is positive.  Hence, there exists a
constant $N$ such that for all $n\geq N$, $f(n_1) \geq 0$.  This means
that our algorithm runs in time $O(n^5)$. \bbox

\begin{theorem}
  \label{t:Strigraph}
  A maximum weighted strong stable set of a trigraph $T$ in $\cal F$ with no
  balanced skew-partition can be computed in time $O(n^5)$.
\end{theorem}
\Proof Run the algorithm from \ref{mainAlg} for $(T, \emptyset,
\emptyset)$.  \bbox

\begin{theorem}
  \label{th:alphaG}
  A maximum weighted stable set of a Berge graph with no balanced skew-partition can be computed in time $O(n^5)$.
\end{theorem}
\Proof Follows from \ref{t:Strigraph} and the fact that a Berge graph may
be seen as a trigraph from $\cal F$.  \bbox

\section{Coloring perfect graphs with an algorithm for stable sets}
\label{sec:color}

Gr\"ostchel, Lov\'asz and Schrijver~\cite{gls:color} proved that the
ellipsoid method yields a polynomial time algorithm that optimally
colors any input perfect graph.  However, so far, no purely
combinatorial method is known.  But, one is known (also due to
Gr\"otschel, Lov\'asz and Schrijver), under the assumption that a
subroutine for computing a maximum stable set is available.  The goal
of this section is to present this algorithm, because it is hard to
extract it from the deeper material that surrounds it in
\cite{gls:color} or~\cite{KrSe:colorP}.

In what follows, $n$ denotes the number of vertices of the graph under
consideration.  We suppose that ${\cal C}$ is a subclass of perfect
graphs, and there is an $O(n^k)$ algorithm ${\cal A}$ that computes a
maximum weighted stable set and a maximum weighted clique for any
input graph in ${\cal C}$.

\begin{theorem}[Lov\'asz \cite{lovasz:nh}]
  \label{th:lovasz}
  A graph is perfect if and only if its complement is perfect.
\end{theorem}

\begin{lemma}
  \label{l:compS}
  There is an algorithm with the following specification:
  \begin{description}
  \item[Input: ] A graph $G$ in ${\cal C}$, and a sequence
    $K_1, \dots, K_t$ of maximum cliques of $G$ where $t\leq n$.
  \item[Output: ] A stable set of $G$ that intersects each $K_i$,
    $i=1, \dots, t$. 
  \item[Running time: ] ${O}(n^k)$
  \end{description} 
\end{lemma}

\Proof
  By $\omega(G)$ we mean here the maximum \emph{cardinality} of a
  clique in~$G$.  Give to each vertex $v$ the weight $y_v= |\{ i; v
  \in K_i \}|$.  Note that this weight is possibly zero.  With
  Algorithm ${\cal A}$, compute a maximum weighted stable set $S$ of~$G$.

  Let us consider the graph $G'$ obtained from $G$ by replicating
  $y_v$ times each vertex $v$. So each vertex $v$ in $G$ becomes a
  stable set $Y_v$ of size $y_v$ in $G'$ and between two such stable sets
  $Y_u$, $Y_v$ there are all possible edges if $uv\in E(G)$ and
  no edges otherwise. Note that vertices of weight zero in $G$
  are not in $V(G')$. Note also that $G'$ may fail to be in ${\cal C}$,
  but it is easily seen to be perfect.  By replicating $y_v$ times
  each vertex $v$ of $S$, we obtain a stable set $S'$ of $G'$ of
  maximum cardinality.

  By construction, $V(G')$ can be partitioned into $t$ cliques of size
  $\omega (G)$ that form an optimal coloring of $\overline{G'}$
  because $\omega(G') = \omega(G)$.  Since by Theorem~\ref{th:lovasz}
  $\overline{G'}$ is perfect, $|S'|=t$.  So, in $G$, $S$ intersects
  every $K_i$, $i \in \{ 1, \ldots ,t\}$.
\bbox

\begin{theorem}
  \label{th:color}
  There exists an algorithm of complexity $O(n^{k+2})$ whose input is
  a graph from ${\cal C}$ and whose output is an optimal coloring of $G$.
\end{theorem}

\Proof
  We only need to show how to find a stable set $S$ intersecting all
  maximum cliques of $G$, since we can apply recursion to $G \setminus
  S$ (by giving weight~0 to vertices of $S$).  Start with $t=0$. At
  each iteration, we have a list of $t$ maximum cliques $K_1, \ldots,
  K_t$ and we compute by the algorithm in Lemma~\ref{l:compS} a stable
  set $S$ that intersects every $K_i$, $i \in \{ 1, \ldots ,t \}$.  If
  $\omega (G \setminus S) < \omega (G)$ then $S$ intersects every
  maximum clique, otherwise we can compute a maximum clique $K_{t+1}$
  of $G \setminus S$ (by giving weight~0 to vertices of~$S$).  This
  will eventually find the desired stable set, the only problem being the
  number of iterations.  We show that this number is bounded by $n$.

  Let $M_t$ be the incidence matrix of the cliques $K_1, \dots, K_t$.
  So the columns of $M_t$ correspond to the vertices of $G$ and each
  row is a clique (we see $K_i$ as row vector).  We prove by induction
  that the rows of $M_t$ are independent.  So, we assume that the rows
  of $M_t$ are independent and prove that this holds again for $M_{t+1}$.

  The incidence vector $x$ of $S$ is a solution to $M_tx = \mathbf{1}$
  but not to $M_{t+1}x = \mathbf{1}$.  If the rows of $M_{t+1}$
  are not independent, we have $K_{t+1} = \lambda_1 K_1 + \cdots +
  \lambda_t K_t$.  Multiplying by $x$, we obtain $K_{t+1}x = \lambda_1
  + \cdots + \lambda_t \neq 1$.  Multiplying by $\mathbf{1}$, we
  obtain $\omega = K_{t+1}\mathbf{1} = \lambda_1 \omega + \cdots +
  \lambda_t \omega$, so $\lambda_1 + \cdots + \lambda_t = 1$, a
  contradiction.

  So the matrices $M_1, M_2, \dots$ cannot have more than $n$
  rows. Hence, there are at most $|V(G)|$ iterations.
\bbox

\subsection*{Proof of Theorem~\ref{th:colorM}}

\Proof An $O(n^5)$ time algorithm exists for the maximum weighted
stable set  by~\ref{th:alphaG}, so an $O(n^7)$ time coloring
algorithm for the same class exists by~\ref{th:color}.  \bbox

\section{Extreme decomposition}
\label{sec:ext}

In this section, we prove that non-basic trigraphs in our class
actually have extreme decompositions. They are decompositions whose
one block of decomposition is basic. Note that this is non-trivial in
general, since in~\cite{nicolas.kristina:2-join} an example
is given, showing that Berge graphs in general do not necessarily have
extreme $2$-joins.  Extreme decompositions are sometimes very useful
for proofs by induction.

In fact, we are not able to prove that any trigraph in our class has
an extreme $2$-join or complement $2$-join; to prove such a statement,
we have to include a new decomposition, the homogeneous pairs, in our
set of decompositions.  Interestingly this decomposition is not new,
it has been used in several variants of Theorem~\ref{noMjointri}.

A {\em proper homogeneous pair} of a trigraph $T$ is a pair of
disjoint nonempty subsets $(A,B)$ of $V(T)$, such that if $A_1,A_2$
denote respectively the sets of all strongly $A$-complete and strongly
$A$-anticomplete vertices and $B_1,B_2$ are defined similarly, then:
\begin{itemize}
\item $|A|>1$ and $|B|>1$;
\item $A_1 \cup A_2=B_1 \cup B_2 = V(T) \setminus (A \cup B)$
(and in particular every vertex in $A$ has a neighbor and an 
antineighbor in $B$ and vice versa); and 
\item the four sets $A_1 \cap B_1$, $A_1 \cap B_2$, $A_2 \cap B_1$, 
$A_2 \cap B_2$ are all nonempty.
\end{itemize}

In these circumstances, we say that $(A, B, A_1 \cap B_2, A_2 \cap B_1,
A_1 \cap B_1, A_2 \cap B_2)$ is a \emph{split} of the homogeneous
pair.

A way to prove the existence of an extreme decomposition is to
consider a ``side'' of a decomposition and to minimize it, to obtain
what we call an \emph{end}.  But for homogeneous pairs, the two sides
(which are $A \cup B$ and $V(T) \setminus (A \cup B)$ with our usual
notation) are not as symmetric as the two sides of a $2$-join, so we
have to decide which side is to be minimized.  We decide to minimize
the side $A \cup B$.  To make all this formal, we therefore have to
distinguish between a \emph{fragment}, which is any side of any
decomposition, and a \emph{proper fragment} which is a side to be
minimized, and therefore cannot be the side $V(T) \setminus (A \cup
B)$ of a homogeneous pair.  All definitions are formally given below.

First we modify our definition of a fragment to include
homogeneous pairs. From here on, A set $X\subseteq V(T)$ is a
\emph{fragment} of a trigraph $T$ if one of the following holds:

\begin{enumerate}
\item\label{i:2J} $(X,V(T)\setminus X)$ is a proper $2$-join of $T$;
\item\label{i:C2J} $(X,V(T)\setminus X)$ is a proper complement $2$-join of $T$;
\item there exists a proper homogeneous pair $(A,B)$ of $T$ such that
$X = A\cup B$ or $X = V(T)\setminus (A\cup B$). 
\end{enumerate}

A set $X\subseteq V(T)$ is a \emph{proper fragment} of a trigraph $T$ if
one of the following holds:
\begin{enumerate}
\item\label{i:2J} $(X,V(T)\setminus X)$ is a proper $2$-join of $T$;
\item\label{i:C2J} $(X,V(T)\setminus X)$ is a proper complement $2$-join of $T$;
\item\label{i:HP} there exists a proper homogeneous pair $(A,B)$ of $T$ such that
$X = A\cup B$.
\end{enumerate}

An \emph{end} of $T$ is a proper fragment $X$ of $T$ such that no
proper induced subtrigraph of $X$ is a proper fragment of $T$.

Note that a proper fragment of $T$ is a proper fragment of
$\overline{T}$, and an end of $T$ is an end of
$\overline{T}$. Moreover a fragment in $T$ is still a fragment in
$\overline{T}$. We have already defined the blocks of decomposition of
a $2$-join or complement-$2$-join.  We now define the blocks of
decomposition of a homogeneous pair.

If $X = A\cup B$ where $(A,B,C,D,E,F)$ is a split of a proper
homogeneous pair $(A,B)$ of $T$, then we build the block of
decomposition as follows. We start with $T | (A\cup B)$. We then add
two new \emph{marker vertices} $c$ and $d$ such that $c$ is strongly
complete to $A$, $d$ is strongly complete to $B$, $cd$ is a
switchable pair, and there are no other edges between $\{c,d\}$ and $A
\cup B$.  Again, $\{c, d\}$ is called the \emph{marker component of~$T_X$}.

If $X = C \cup D \cup E \cup F$ where $(A,B,C,D,E,F)$ is a
split of a proper homogeneous pair $(A,B)$ of $T$, then we build the
block of decomposition $T_{X}$ with respect to $X$ as follows. We
start with $T|X$. We then add two new \emph{marker vertices} $a$ and
$b$ such that $a$ is strongly complete to $C\cup E$, $b$ is strongly
complete to $D\cup E$, $ab$ is a switchable pair, and there are no
other edges between $\{a,b\}$ and $C \cup D \cup E \cup F$.  Again,
$\{a, b\}$ is called the \emph{marker component of $T_X$}.

\begin{theorem}
 \label{l:stayBergeExt}
 If $X$ is a fragment of a trigraph $T$ from $\mathcal F$ with no
 balanced skew-partition, then $T_X$ is a trigraph from~$\mathcal F$.
\end{theorem}

\Proof From the definition of $T_X$, it is clear that every vertex of
$T_X$ is in at most one switchable pair, or is heavy, or is light.
So, to prove that $T_X \in {\mathcal F}$, it remains only to prove
that $T_X$ is Berge.

If the fragment come from a $2$-join or the complement of a $2$-join,
we have the result by~\ref{l:stayBerge}.

If $X = A \cup B$ and $(A, B)$ is a proper homogeneous pair of $T$,
then let $H$ be a hole or an antihole in $T_X$. Passing to the
complement if necessary, we may assume that $H$ is a hole. If it
contains the two markers $c, d$, it must be a cycle on four vertices,
or it must contain two strong neighbors of $c$ in $A$, and two strong
neighbors of $d$ in $B$, so $H$ has length~6. Hence, we may assume
that $H$ contains at most one of $c, d$, so a hole of the same length
in $T$ is obtained by possibly replacing $c$ or $d$ by some vertex of
$C$ or $D$. Hence, $H$ has even length.

If there exists a proper homogeneous pair $(A,B)$ of $T$ such that $X
= V(T) \setminus (A\cup B)$, then since every vertex of $A$ has a
neighbor and an antineighbor in $B$, we see that every realization of
$T_X$ is an induced subgraph of some realization of $T$. It follows
that $T_X$ is Berge.  \bbox

\begin{theorem}\label{stabExt}
  If $X$ is a fragment of a trigraph $T$ from $\mathcal{F}$ with no
  balanced skew-partition, then the block of decomposition $T_X$ has
  no balanced skew-partition.
\end{theorem}

\Proof To prove this, we suppose that $T_X$ has a balanced skew-partition $(A',B')$ with a split $(A_1',A_2',B_1',B_2')$.  From this,
we find a skew-partition in $T$.  Then we use~\ref{onepair} to prove
the existence of a \emph{balanced} skew-partition in $T$.  This gives
a contradiction that proves the theorem.

If the fragment come from a $2$-join or the complement of a $2$-join,
we have the result by~\ref{stab}.

If $X = A \cup B$ and $(A, B)$ is a homogeneous
pair of $T$, then  let $(A,B,C,D,E,F)$ be a split of $(A,B)$.
Because $cd$ is a switchable pair, the markers $c$ and $d$ have no
common neighbor and $cd$ dominates $T_X$, there is up to symmetry only
one case: $c\in A'_1$ and $d\in B'_1$. Since $B_2'$ is complete to
$d$, and $A_2'$ is anticomplete to $c$, it follows that $A_2',B_2'
\subseteq B$.

Now $(A_1'\setminus\{c\}\cup C\cup F, A_2', B_1'\setminus\{d\}\cup
D\cup E, B_2')$ is a split of a skew-partition  in
$T$. The pair $(A_2', B_2')$ is balanced in $T$ because it is balanced
in $T_X$.  Hence, by~\ref{onepair}, $T$ admits a balanced skew-partition, a contradiction.

If 
$X=V(T)\setminus (A\cup B)$ and $(A, B)$ is a proper homogeneous pair
of $T$, then  let $(A,B,C,D,E,F)$ be a split of $(A,B)$.  
Because $ab$ is a switchable pair we may assume, using symmetry and
complementation that $a\in A_1'$ and $b\in A_1'\cup B_1'$. If $b\in
A_1'$, then $(A\cup B\cup A_1'\setminus \{a, b\}, A_2' , B_1' , B_2')$
is a split of a skew-partition in $T$, and if $b\in B_1'$ , then
$(A\cup A_1'\setminus \{a\}, A_2' , B\cup B_1'\setminus \{b\}, B_2')$
is a split of a skew-partition in $T$.  In both cases, the pair
$(A_2', B_2')$ is balanced in $T$ because it is balanced in $T_X$.
Hence, by~\ref{onepair}, $T$ admits a balanced skew-partition, a
contradiction. \bbox

\begin{theorem} 
 \label{l:extreme}
 If $X$ is an end of a trigraph $T$ from $\mathcal{F}$ with no balanced
 skew-partition, then the block of decomposition $T_X$ is basic.
\end{theorem}

\Proof Let $T$ be a trigraph from $\mathcal{F}$ with no balanced
skew-partition and $X$ an end of $T$. By~\ref{l:stayBergeExt}, we know
that $T_X\in \mathcal{F}$ and by~\ref{stabExt}, we know that $T_X$ has no
balanced skew-partition. By~\ref{structure}, it is enough to show that
$T_X$ has no proper $2$-join and no proper complement $2$-join.

Passing to the complement if necessary, we may assume that one of the
following three statements hold:
\begin{itemize}
\item $X=A\cup B$ and $(A,B)$ is a proper homogeneous pair of $T$;
\item $(X,V(T)\setminus X)$ is a proper even $2$-join of $T$;
\item $(X,V(T)\setminus X)$ is a proper odd $2$-join of $T$.
\end{itemize}

\noindent {\bf Case 1:} $X = A\cup B$ where $(A,B)$ is a proper
homogeneous pair of $T$. Let $(A,B,C,D,E,F)$ be a split of $(A,B)$.

Suppose $T_X$ admits a proper $2$-join $(X_1,X_2)$. Let
$(A_1,B_1,C_1,A_2,B_2,C_2)$ be a split of $(X_1,X_2)$. Because $cd$
is a switchable pair we may assume that $c,d$ are both in $X_2$. As
$\{c,d\}$ strongly dominates $T_X$ we may assume that $c\in A_2$ and
$d\in B_2$, so $C_1=\emptyset$. Since $c$ is strongly complete
to $A$, $A_1\subseteq A$, and analogously $B_1\subseteq B$.
By~\ref{stabExt} and~\ref{2joinform}, $|A_1|\geq 2$ and $|B_1|\geq 2$, and because
$C_1=\emptyset$, every vertex from $A_1$ has a neighbor and an
antineighbor in $B_1$ and vice versa. Now $(A_1, B_1, C\cup
A_2\setminus \{c\}, D \cup B_2 \setminus \{d\}, E, F\cup C_2)$ is
a split of a proper homogeneous pair of $T$. Because $|X_2|\geq 3$,
$A_1\cup B_1$ is strictly included in $A\cup B$, a contradiction.

Because $A\cup B$ is also a homogeneous pair of $\overline{T}$, by the
same argument as above, $T_X$ cannot admit a proper complement $2$-join.

\noindent {\bf Case 2:} $(X,V(T)\setminus X)$ is a proper even $2$-join
$(X_1, X_2)$ of $T$, where $X=X_1$. Let $(A_1,B_1,C_1,A_2,B_2,C_2)$ be a split of
$(X,V(T)\setminus X)$.

Suppose that $T_X$ admits a proper $2$-join $(X'_1,X'_2)$. Let
$(A'_1,B'_1,C'_1,A'_2,B'_2,C'_2)$ be a split of $(X'_1,X'_2)$. Since
$ac$ and $bc$ are switchable pairs, we may assume that $a,b,c \in
X'_2$. Now we claim that $(X'_1, V(T)\setminus X'_1)$ is a proper
$2$-join of $T$ and $X'_1$ is strictly included in $X$, which gives a
contradiction. Note that because of the definition of a $2$-join and
the fact that $c$ has no strong neighbor, $X'_2$ cannot only be $\{a,
b, c\}$ and hence, $X'_1$ is strictly included in $X$.  Since $c$ has
no strong neighbor, we have $c\in C'_2$. Since $a$ and $b$ have no
common strong neighbor in $T_{X_1}$, there are up to symmetry three
cases: either $a\in A_2'$, $b\in B_2'$, or $a\in A_2'$, $b\in C_2'$,
or $a,b\in C_2'$.

If $a\in A'_2$ and $b\in B'_2$, then $(A'_1, B'_1, C'_1, A_2 \cup
A'_2\setminus \{a\}, B_2 \cup B'_2\setminus \{b\}, C_2 \cup
C'_2\setminus\{c\})$ is a split of a  $2$-join of $T$.

If $a\in A'_2$ and $b\in C'_2$, then $(A'_1,B'_1,C'_1, A_2 \cup
A'_2\setminus \{a\}, B'_2, B_2\cup C_2 \cup C'_2\setminus\{b,c\})$ is
a split of a $2$-join of $T$.

If $a\in C'_2$ and $b\in C'_2$, then $(A'_1, B'_1, C'_1, A'_2, B'_2,
X_2 \cup C'_2\setminus\{a,b,c\})$ is a split of a  $2$-join of
$T$.

By~\ref{stabExt} and~\ref{2joinform} each of these $2$-joins is proper, and we have a
contradiction.

Suppose $T_X$ admits a proper complement $2$-join $(X'_1,X'_2)$. Because
$c$ has no strong neighbor we get a contradiction.

\noindent {\bf Case 3:} $(X,V(T)\setminus X)$ is a proper odd $2$-join
$(X_1, X_2)$ of $T$, where $X=X_1$. Let $(A_1,B_1,C_1,A_2,B_2,C_2)$ be a split of
$(X,V(T)\setminus X)$.

Suppose $T_X$ admits a proper $2$-join $(X'_1,X'_2)$. Let
$(A'_1,B'_1,C'_1,A'_2,B'_2,C'_2)$ be a split of $(X'_1,X'_2)$. Since
$ab$ is a switchable pair, we may assume that $a,b\in X'_2$. Now we
claim that $(X'_1, V(T)\setminus X'_1)$ is a proper $2$-join of $T$,
obtaining a contradiction, because $X'_2$ cannot be only $\{a,b\}$ (by
the definition of a $2$-join), so $X'_1$ is strictly included in $X$.
Because $a$ and $b$ have no common strong neighbor in $T_{X_1}$ there
are up to symmetry three cases: either $a\in A_2'$, $b\in B_2'$, or
$a\in A_2'$, $b\in C_2'$, or $a,b\in C_2'$.

If $a\in A'_2$ and $b\in B'_2$, then $(A'_1,B'_1,C'_1, A_2 \cup
A'_2\setminus \{a\}, B_2 \cup B'_2\setminus \{b\}, C_2 \cup C'_2)$ is
a split of a  $2$-join of $T$.

If $a\in A'_2$ and $b\in C'_2$, then $(A'_1,B'_1,C'_1, A_2 \cup
A'_2\setminus \{a\}, B'_2, B_2\cup C_2 \cup C'_2\setminus\{b\})$ is a
split of a  $2$-join of $T$.

If $a\in C'_2$ and $b\in C'_2$, then $(A'_1,B'_1,C'_1,A'_2, B'_2, X_2
\cup C'_2\setminus\{a,b\})$ is a split of a  $2$-join of $T$.

By~\ref{stabExt} and~\ref{2joinform} each of these $2$-joins is proper, and we have a
contradiction.

Suppose $T_X$ admits a proper complement $2$-join $(X'_1,X'_2)$. Let
$(A'_1,B'_1,C'_1,A'_2,B'_2,C'_2)$ be a split of $(X'_1,X'_2)$.
Because $ab$ is a switchable pair, we may assume that $a,b\in X'_2$.
Because $a$ and $b$ have no common strong neighbor we may assume that
$a\in A'_2$, $b\in B'_2$ and $C'_1=\emptyset$. If $C_2$ and $C'_2$ are
not empty, then $(A'_1,B'_1,B_2 \cup B'_2\setminus\{b\}, A_2 \cup
A'_2\setminus \{a\}, C'_2, C_2)$ is a split of a proper homogeneous
pair of $T$ and $A'_1\cup B'_1$ is strictly included in $X$, a
contradiction (note that by~\ref{stabExt} and~\ref{2joinform},
$|A'_1|\geq 2$, $|B'_1|\geq 2$, and each vertex from $A'_1$ has a
neighbor and an antineighbor in $B'_1$ and vice versa). If $C_2$ is
not empty and $C'_2$ is empty, then $(A'_1,B'_1,\emptyset, B_2 \cup
B'_2\setminus\{b\}, A_2 \cup A'_2\setminus\{a\}, C_2)$ is a split of a
proper $2$-join of $T$ (the $2$-join is proper by~\ref{stabExt}
and~\ref{2joinform}). If $C_2$ is empty, then $(A'_1,B'_1,\emptyset,
A_2 \cup A'_2\setminus\{a\}, B_2 \cup B'_2\setminus\{b\}, C'_2)$ is a
split of a proper complement $2$-join of $T$ (again, it is proper
by~\ref{stabExt} and~\ref{2joinform}).  \bbox

\section{Means to an end}
\label{sec:end}

The goal of this section is to describe a polynomial time algorithm
that outputs an end (defined in Section~\ref{sec:ext}) of an input
trigraph (if any).  To do so, one may rely on existing algorithms for
detecting $2$-joins and homogeneous pairs.  The fastest one is
in~\cite{ChHaTrVu:2-join} for $2$-joins and~\cite{HaMaMo:HP} for
homogeneous pairs.  But there are several problems with this approach.
First, all the classical algorithms work for graphs, not for
trigraphs.  They are easy to convert into algorithms for trigraphs,
but it is hard to get convinced by that without going through all the
algorithms.  Worse, most of the algorithms output a fragment, not an
end.  In fact, for the $2$-join, an algorithm
from~\cite{ChHaTrVu:2-join} does output a minimal set $X$ such that
$(X, V(G)\setminus X)$ is a $2$-join, but there still could be a
homogeneous pair inside $X$.  So, we prefer to write our own
algorithm, even if most ideas are from existing work.

Our algorithm looks for a proper fragment $X$.  Because all the
technical requirements in the definitions of $2$-joins and homogeneous
pairs are a bit messy, we introduce a new notion.  A \emph{weak
  fragment} of a trigraph $T$ is a set $X\subseteq V(T)$ such that
there exist disjoint sets $A_1$, $B_1$, $C_1$, $D_1$, $A_2$, $B_2$,
$C_2$, $D_2$ satisfying:

\begin{itemize}
\item $X = A_1 \cup B_1 \cup C_1 \cup D_1$;
\item  $V(T) \setminus X= A_2 \cup B_2 \cup C_2 \cup D_2$;
\item $A_1$ is strongly complete to $A_{2} \cup D_{2}$ and strongly
  anticomplete to $B_2 \cup C_2$;
\item $B_1$ is strongly complete to $B_{2}\cup D_{2}$ and strongly
 anticomplete to $A_2 \cup C_2$;
\item $C_1$ is strongly anticomplete to $A_2 \cup B_{2}\cup C_{2}$;
\item $D_1$ is strongly complete to $A_{2}\cup B_{2}\cup D_{2}$;
\item $|X|\geq 4$ and $|V(T) \setminus X| \geq 4$;
\item $|A_i| \geq 1$ and $|B_i| \geq 1$, $i=1, 2$;
\item and at least one of the following statement:

\begin{itemize}
\item $C_1=D_1 = \emptyset$, $C_2 \neq \emptyset$, and $D_2 \neq \emptyset$,  or
\item $D_1=D_2=\emptyset$, or
\item $C_1=C_2 = \emptyset$.
\end{itemize}
\end{itemize}

In these circumstances, we say that $(A_1, B_1, C_1, D_1, A_2, B_2,
C_2, D_2)$ is a \emph{split} for $X$.
Given a weak fragment we say it is of type {\em homogeneous pair}
if $C_1=D_1=\emptyset$, $C_2 \neq \emptyset$, and $D_2 \neq \emptyset$, of 
type {\em $2$-join} if $D_1=D_2=\emptyset$, and
of type {complement $2$-join} if $C_1=C_2=\emptyset$. 
Note that a weak fragment may be simultaneously a $2$-join fragment and
a complement $2$-join fragment (when $C_1 = D_1 = C_2 = D_2 =
\emptyset$). 

\begin{theorem}
\label{weakstruct}
If $T$ is a trigraph from $\mathcal{F}$ with no balanced
skew-partition, then $X$ is a weak fragment of $T$ if and only if $X$
is a proper fragment of $T$.
\end{theorem}

\Proof If $X$ is a proper fragment, then it is clearly a weak fragment (the
conditions $|X| \geq 4$ and $|V(T) \setminus X| \geq 4$ are satisfied
when $X$ is a side of a $2$-join by~\ref{2joinform}).  Let us prove the
converse. Let $X$ be a weak fragment, and let 
$(A_1,B_1,C_1,D_1, A_2,B_2,C_2,D_2)$ be a split for $X$. If $X$
is of type $2$-join  or complement $2$-join, then  it is proper
by~\ref{2joinform}. Thus we may assume that $X$ is
of type homogeneous pair, and so
$C_1=D_1=\emptyset$, $C_2 \neq \emptyset$, and $D_2 \neq \emptyset$.
Since all 4 sets $A_1,A_2,B_1,B_2$ are non-empty, it remains to check the
following:
\begin{itemize}
\item[(i)] Every vertex of $A_1 (B_1)$ has a neighbor and antineighbor
in $B_1(A_1)$. 
\item[(ii)] $|A_1|>1$ and $|B_1|>1$.
\end{itemize}

Suppose (i) does not hold. By passing to $\overline{T}$ if necessary,
we may assume that some $v \in A_1$ is strongly complete to
$B_1$. Since $\{v\} \cup B_1 \cup A_2 \cup D_2$ is not a star cutset
in $T$ by~\ref{starcutset}, it follows that $A_1=\{v\}$. Now every
vertex of $B_1$ is strongly complete to $A_1$, and so, by the same
argument, $|B_1|=1$, contradicting the assumption that $|X|\geq
4$. Therefore (i) holds.

To prove (ii) assume that $|A_1|=1$. Since $|X|\geq 4$, it follows that $|B_1|\geq 3$.
By (i)  every vertex
of $B_1$ is semi-adjacent to the unique vertex of $A_1$, which is impossible
since $|B_1|\geq 3$ and $T \in \mathcal{F}$. Therefore (ii) holds.
\bbox

A $4$-tuple $(a_1, b_1, a_2, b_2)$ of vertices from a trigraph $T$ is
\emph{proper} if:

\begin{itemize}
\item $a_1$, $b_1$, $a_2$, $b_2$ are pairwise distinct;
\item $a_1a_2, b_1b_2 \in \eta(T)$;
\item $a_1b_2, b_1a_2 \in \nu(T)$.
\end{itemize}

A proper $4$-tuple $(a_1, b_1, a_2, b_2)$ is \emph{compatible} with a
weak fragment $X$ if there is a split
$(A_1,B_1,C_1,D_1,A_2,B_2,C_2,D_2)$ for $X$ such that $a_1\in A_1$,
$b_1\in B_1$, $a_2\in A_2$ and $b_2\in B_2$.

We use the following notation. When $x$ is a vertex of a trigraph $T$,
$N(x)$ denotes the set of the neighbors of $x$, $\overline{N}(x)$
denotes the set of the antineighbors of $x$, $\eta(x)$ the set of the
strong neighbors of $x$, and $\sigma(x)$ the set of vertices $v$ such
that $xv \in \sigma(T)$.

\begin{theorem}\label{l:forcing}
  Let $T$ be a trigraph and $Z = (a_1, b_1, a_2, b_2)$ a proper
  $4$-tuple of $T$.  There is an $O(n^2)$ time algorithm that given a
  set $R_0 \subseteq V(T)$ of size at least $4$ such that $Z \cap R_0
  = \{a_1, b_1\}$, outputs a weak fragment $X$ compatible with $Z$ and
  such that $R_0 \subseteq X$, or outputs the true statement ``There
  exists no weak fragment $X$ compatible with $Z$ and such that $R_0
  \subseteq X$''.

  Moreover, when $X$ is outputted, it is minimal with respect to
  these properties, meaning that $X \subset X'$ for every weak
  fragment $X'$ satisfying the properties.
\end{theorem}

\Proof
\begin{table}{\small\label{algoM}
\begin{description}
\item{\bf Input:} $R_0$ a  set of vertices of a trigraph $T$ and a proper $4$-tuple $Z = (a_1, b_1, a_2, b_2)$ such
  that $a_{1}, b_{1} \in R_0$ and $a_{2}, b_{2} \notin R_0$.

\item{\bf Initialization:}

$R \leftarrow R_0$; $S \leftarrow V(T) \setminus R_0$; 
$A \leftarrow \eta(a_{1})\cap S$;  $B \leftarrow \eta(b_{1})\cap S$;\\
$\State \leftarrow \text{Unknown}$;

Vertices $a_{1}, b_{1}, a_{2}, b_{2}$ are left unmarked. For the other vertices of $T$:

\rule{1em}{0ex}$\Mark(x) \leftarrow \alpha\beta$ for every vertex $x \in  \eta(a_{2})\cap \eta(b_{2}) $;

\rule{1em}{0ex}$\Mark(x) \leftarrow \alpha$ for every vertex $x \in  \eta(a_{2}) \setminus \eta(b_{2}) $;

\rule{1em}{0ex}$\Mark(x) \leftarrow \beta$ for every vertex $x \in  \eta(b_{2}) \setminus \eta(a_{2}) $;

\rule{1em}{0ex}Every other vertex of $T$ is marked by $\varepsilon$;

\rule{1em}{0ex}$\Move(\sigma(a_1) \cap S)$; $\Move(\sigma(b_1) \cap S)$;

\item{\bf Main loop:}

\textbf{While} there exists a vertex $x \in R$ marked

\textbf{Do} $\Explore(x); \Unmark(x);$

\item{\bf Function Explore(x):}

\rule{1em}{0ex}\textbf{If} 
$\Mark(x)=\alpha\beta$ \textbf{and} $\State=\text{Unknown}$ \textbf{then}

\rule{2em}{0ex}$\State\leftarrow \overline{\text{$2$-join}}$, $\Move(S
\setminus (A
\cup B))$;

\rule{1em}{0ex}\textbf{If} $\Mark(x)=\alpha\beta$ \textbf{and}
$\State=\overline{\text{$2$-join}}$ \textbf{then}  
$\Move( \overline{N}(x)\cap S)$;

\rule{1em}{0ex}\textbf{If} $\Mark(x)=\alpha\beta$ \textbf{and}
$\State=\text{$2$-join}$ \textbf{then}\\
\rule{2em}{0ex}{\textbf{Output} {\it No weak fragment is found}}, 
\textbf{Stop};

\rule{1em}{0ex}\textbf{If} $\Mark(x)=\alpha$ \textbf{then}  
$\Move(A \Delta (\eta(x)\cap S))$, $\Move(\sigma(x)\cap S)$;

\rule{1em}{0ex}\textbf{If} $\Mark(x)=\beta$ \textbf{then}  
$\Move(B \Delta (\eta(x)\cap S))$, $\Move(\sigma(x)\cap S)$;

\rule{1em}{0ex}\textbf{If} $\Mark(x)=\varepsilon$ \textbf{and}
$\State=\text{Unknown}$ \textbf{then} 

\rule{2em}{0ex}$\State\leftarrow \text{$2$-join}$, $\Move(A\cap B)$;

\rule{1em}{0ex}\textbf{If} $\Mark(x)=\varepsilon$ \textbf{and} $\State=\text{$2$-join}$ \textbf{then} 
$\Move(N(x)\cap S)$;

\rule{1em}{0ex}\textbf{If} $\Mark(x)=\varepsilon$ \textbf{and} $\State=\overline{\text{$2$-join}}$ \textbf{then} \\
\rule{2em}{0ex}{\textbf{Output} {\it No weak fragment is found}}, 
\textbf{Stop};

\item{\bf Function Move(Y):}

\rule{1em}{0ex}{\it This function just moves a subset $Y \subset S$ from $S$ to $R$. }

\rule{1em}{0ex}\textbf{If} $Y\cap \{a_2,b_2\}\not=\emptyset$ \textbf{then} \\
\rule{2em}{0ex}{\textbf{Output} {\it No weak fragment is found}}, 
\textbf{Stop};

\rule{1em}{0ex}{$R \leftarrow R \cup Y$;  $ A\leftarrow A \setminus Y$;   $ B\leftarrow B \setminus Y$;  $ S\leftarrow S \setminus Y$;}
\end{description}

\caption{Procedure used in Theorem~\ref{l:forcing}\label{algoM}}}
\end{table}
We use the procedure described in Table~\ref{algoM}.  It
tries to build a weak fragment $R$, starting with $R = R_0$ and $S=V(T)\setminus R_0$.  Then,
several forcing rules are implemented, stating that some sets of
vertices must be moved from $S$ to $R$.  The variable ``State''
contains the type of the weak fragment that is being considered.  At the
beginning, it is ``Unknown''.   The following
properties are easily checked to be invariant during all the
execution of the procedure (meaning that they are satisfied after each
call to Explore):

\begin{itemize}
\item $R$ and $S$ form a partition of $V(T)$, $R_0\subseteq  R$ and
  $a_2, b_2 \in S$.

\item For all unmarked $v \in R$, and all $u\in S$, $uv$ is not a
  switchable pair.  

\item All unmarked vertices belonging to $R \cap (\eta(a_{2})
  \setminus \eta(b_2))$ have the same neighborhood in $S$, namely $A$
  (and $A$ is a strong neighborhood).

\item All unmarked vertices belonging to $R\cap (\eta(b_{2}) \setminus
  \eta(a_2))$ have the same neighborhood in $S$, namely $B$ (and $B$
  is a strong neighborhood).

\item All unmarked vertices belonging to $R\cap (\eta(b_{2}) \cap
  \eta(a_2))$ have the same neighborhood in $S$, namely $A \cup B$.

\item All unmarked  vertices belonging to $R$ not adjacent to $a_{2}$ 
nor $b_{2}$ are strongly anticomplete to $S$.

\item For every weak fragment $X$ such that $R_0 \subseteq X$ and
  $a_2,b_2 \in V(T)\setminus X$, we have that $R \subseteq X$ and
  $V(T)\setminus X \subseteq S$.
\end{itemize}

By the last item all moves from $S$ to $R$ are necessary.  Hence, if
some vertex in $R$ is strongly adjacent to $a_2$ and $b_2$, any weak
fragment compatible with $Z$ that contains $R$ must be a complement
$2$-join fragment.  This is why the variable State is assigned value
$\overline{\text{$2$-join}}$ and all vertices of $S\setminus (A \cup
B)$ are moved to $R$.  Similarly, if some vertex in $R$ is strongly
antiadjacent to $a_2$ and $b_2$, any weak fragment compatible with $Z$
that contains $R$ must be a $2$-join fragment.  This is why the variable
State is assigned value ${\text{$2$-join}}$ and all vertices of $A
\cap B$ are moved to $R$.

When $\State = \overline{\text{$2$-join}}$ and a vertex in $R$ is
discovered to be strongly antiadjacent to $a_2$ and $b_2$, there is a
contradiction with the definition of the complement $2$-join, so the
algorithm must stop.  When $\State = {\text{$2$-join}}$ and a vertex
in $R$ is discovered to be strongly adjacent to $a_2$ and $b_2$, there
is a contradiction with the definition of the $2$-join, so the
algorithm must stop.  When the function Move tries to move $a_2$ or
$b_2$ in $R$ (this may happen if some vertex in $R$ is semiadjacent to $a_2$
or $b_2$), then $R$ cannot be contained in any
fragment compatible with $Z$.

If the process does not stop for all the reasons above, then all
vertices of $R$ have been explored and therefore are unmarked.  So, if
$|S|\geq 4$, at the end, $R$, is a weak fragment compatible with $Z$.
More specifically, $(R \cap (\eta(a_2) \setminus \eta(b_2)), R
\cap(\eta(b_2) \setminus \eta(a_2)), R \setminus (\eta(a_2) \cup
\eta(b_2)), R \cap (\eta(a_2) \cap \eta(b_2)), A \setminus B, B
\setminus A, S \setminus (A \cup B), A \cap B )$ is a split for the
weak fragment $R$.

Since all moves from $S$ to $R$ are necessary, the fragment is minimal
as claimed.  This also implies that if $|S| \leq 3$, then no desired
fragment exists, in which case, the algorithm outputs that no weak
fragment exists.

\noindent\textbf{Complexity Issues:} The neighborhood and
antineighborhood of a vertex in $R$ is considered at most once.  So,
globally, the process requires $O(n^2)$ time.  \bbox

\begin{theorem}
  \label{th:detect}
  There exists an $O(n^5)$ time algorithm whose input is a trigraph
  $T$ from $\cal F$ with no balanced skew-partition, and whose output
  is an end $X$ of $T$ (if any such end exists) and the block $T_X$.
\end{theorem}

\Proof Recall that by~\ref{weakstruct}, the weak fragments of $T$ are
its proper fragments.  We first describe an $O(n^8)$ time algorithm,
and then we explain how to speed it up.  We assume that $|V(T)|\geq 8$
for otherwise no proper fragment exists.  For all proper $4$-tuple $Z
= (a_1, a_2, b_1, b_2)$ and for all pairs of vertices $u, v$ of $V(T)
\setminus \{ a_1,a_2,b_1,b_2\}$, we apply~\ref{l:forcing} to $R_0=\{
a_1, b_1, u, v\}$.  This method detects for each $Z$ and each $u, v$ a
proper fragment compatible with $Z$, containing $u, v$, and minimal
with respect to these properties (if any).  Among all these fragments,
we choose one with minimum cardinality, this is an end.  Once the end is
given, it is easy to know the type of decomposition that is used and
to build the corresponding block (in particular, by~\ref{l:par2Join},
one may test by just checking one path whether a $2$-join is odd or
even).  Let us now explain how to speed this up.

We look for $2$-joins and homogeneous pairs separately.  We describe
an $O(n^5)$ time procedure that outputs a $2$-join weak fragment, an
$O(n^5)$ time procedure that outputs a complement $2$-join weak
fragment, and an $O(n^5)$ time procedure that outputs a homogeneous
pair weak fragment.  Each of them outputs a fragment of minimum
cardinality among all fragments of its respective kind.  Hence, a
fragment of minimum cardinality chosen among the three is an end.

Let us first deal with $2$-joins.  A set $\cal Z$ of proper $4$-tuples
is \emph{universal} if for every proper $2$-join with split $(A_1,
B_1, C_1, A_2, B_2, C_2)$, there exists $(a_1, a_2, b_1, b_2) \in
{\cal Z}$ such that $a_1 \in A_1$, $a_2 \in A_2$, $b_1 \in B_1$, $b_2
\in B_2$.  Instead of testing all $4$-tuples as in the $O(n^8)$ time
algorithm above, it is obviously enough to restrict the search to a
universal set of $4$-tuples.  As proved in~\cite{ChHaTrVu:2-join},
there exists an algorithm that generates in time $O(n^2)$ a universal
set of size at most $O(n^2)$ for any input graph.  It is easy to
obtain a similar algorithm for trigraphs.

The next idea for $2$-joins is to apply the method from
Table~\ref{algoM} to $R_0=\{ a_1, b_1, u\}$ for all $u$'s instead of
$R_0=\{ a_1, b_1, u, v\}$ for all $u, v$'s.  As we explain now, this
finds a $2$-join compatible with $Z = (a_1, a_2, b_1, b_2)$ when there
is one.  For suppose $(X_1, X_2)$ is such a $2$-join.  If $X_1$ contains
a vertex $v$ whose neighborhood (in $T$) is different from $\{a_1,
b_1\}$, then by~\ref{2joinform}, $v$ has at least one neighbor in
$V(T) \setminus \{a_1, b_1\}$.  Hence, when the loop considers $u=v$,
the method from Table~\ref{algoM} moves some new vertices in $R$.  So,
at the end, $|R| \geq 4$ and the $2$-join is detected.  So, the method
fails to detect a $2$-join only when $v$ has degree~2 and $a_1 \d v \d
b_1$ is a path while a $2$-join compatible with $Z$ exists, with $v$
in the same side as $a_1, b_1$.  In fact, since all vertices $u$ are
tried, this is a problem only if this failure occurs for every
possible $u$, that is if the $2$-join we look for has one side made of
$a_1$, $b_1$, and a bunch of vertices $u_1, \dots, u_k$ of degree~2
all adjacent to $a_1$ and $b_1$.  But in this case, either one of the
$u_i$'s is strongly complete to $\{a_1, b_1\}$ and it is the center of
star cutset, or all the $u_i$'s are adjacent to at least one of $a_1,
b_1$ by a switchable pair.  In this last case, all the $u_i$'s are
moved to $R$ when we run the method from Table~\ref{algoM}, so the
$2$-join is in fact detected.

Complement $2$-joins are handled by the same method in the complement.

Let us now consider homogeneous pairs.  It is convenient to define
\emph{weak homogeneous pairs} exactly as proper homogeneous pairs,
except that we require that ``$|A|\geq 1$, $|B|\geq 1$ and $|A\cup B|
\geq 3$'' instead of ``$|A|>1$ and $|B|>1$''.  A theorem similar
to~\ref{l:forcing} exists, where the input of the algorithm is a graph
$G$, a triple $(a_1, b_1, a_2) \in V(G)^3$ and a set $R_0\subseteq
V(G)$ that contains $a_1, b_1$ but not $a_2$, and the output is a weak
homogeneous pair $(A, B)$ such that $R_0 \subseteq A \cup B$, $a_1 \in
A$, $b_1\in B$ and $a_2\notin A\cup B$, and such that $a_2$ is
complete to $A$ and anticomplete to $B$, if any such weak homogeneous
pair exits.  As in~\ref{l:forcing}, the running time is $O(n^2)$ and
the weak homogeneous pair is minimal among all possible weak
homogeneous pairs.  This is proved in~\cite{everett.k.r:findingHP}.

As for $2$-joins, we define the notion of a universal set of triples
$(a_1, b_1, a_2)$.  As proved in~\cite{HaMaMo:HP}, there exists an
algorithm that generates in time $O(n^2)$ a universal set of size at
most $O(n^2)$ of triples for any input graph.  It is very easy to
obtain a similar algorithm for trigraphs.  As in the $2$-join case, we
apply the analogue of ~\ref{l:forcing} to all vertices $u$ instead of
all pairs $u, v$.  The only problem is when after the call to the
analogue of~\ref{l:forcing}, we have a weak and non-proper homogeneous
pair (so $|A \cup B| = 3$).  But then, it can be checked that the
trigraph has a star cutset or a star cutset in the complement.  \bbox

\section{Enlarging the class: open questions}
\label{sec:enlarge}

 The class $\cal C$ of Berge graphs for which we are able to compute maximum
 stable sets, namely Berge graphs with no balanced skew-partitions, has a
 strange disease: it is not closed under taking induced subgraphs.
  But from an algorithmic point of view,
 since we are able to do the computations with weights on the
 vertices, we can simulate ``taking an induced subgraph'' by putting
 weight zero on the vertices that we want to delete.

\begin{figure}
  \begin{center}
    \includegraphics{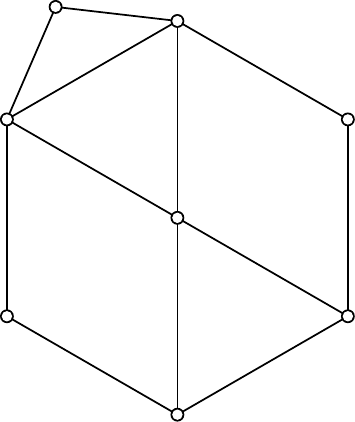}
    \caption{A graph with a balanced skew-partition\label{fig:contrex4}}
  \end{center}
\end{figure}

This suggests that in fact, we work on the more general class ${\cal
  C}'$ of graphs that are  induced subgraphs of some graph in
$\cal C$.  The class ${\cal C}'$ is closed under taking
induced subgraphs so it must be defined by a list of forbidden induced
subgraphs.  We leave the following questions open: what are the
forbidden induced subgraphs for ${\cal C}'$?  One could think that
${\cal C}'$ is in fact the class of all Berge graphs, but it is not
the case as shown by the graph $G$ represented
in Figure~\ref{fig:contrex4}.  The graph $G$ is Berge and admits an obvious
balanced skew-partition.  Moreover,  Robertson, Seymour and Thomas proved
that a Berge graph that contains $G$ as an induced subgraph also
admits a balanced skew-partition, see~\cite{seymour:how}, page~78.
So, $G$ is not in ${\cal C}'$ and $G$ might be the smallest example of
a Berge graph not in ${\cal C}'$.

Here are more questions on $\cal C$ and ${\cal C}'$.
For any graph $G$ in ${\cal C}'$, is there a graph $H$ in
${\cal C}$ whose size is polynomial in the size of $G$ and such
that $G$ is an induced subgraph of $H$?  If yes, or when yes, can we
compute $H$ from $G$ in polynomial time?

\section*{Acknowledgment}

Thanks to Antoine Mamcarz for useful discussions on how to detect an
end of a trigraph. Thanks to Fabien de Montgolfier for having
suggesting to us that the complexity analysis in the proof of
Theorem~\ref{mainAlg} should exist.  The work on this paper began when
the first and last author were visiting LIAFA under the generous
support of Universit\'e Paris~7.

\end{document}